\documentclass{aa}  

\usepackage{graphicx}
\usepackage[varg]{txfonts}
\usepackage{epstopdf}
\usepackage{lipsum}
\usepackage{subcaption}         
\usepackage{lscape}             
\usepackage{placeins}           
\usepackage{amsmath}

\usepackage[colorlinks=true,linkcolor=magenta,citecolor=blue,filecolor=black,urlcolor=cyan,final=true]{hyperref}

\begin{document}

   \title{The ASPIICS solar coronagraph aboard the Proba-3 formation flying mission}

   \subtitle{Scientific objectives and instrument design}

   \author{
   A.~N.~Zhukov\inst{\ref{ROB}}
   \and
   C.~Thizy\inst{\ref{CSL}}
   \and
   D.~Galano\inst{\ref{ESTEC}}
   \and
   B.~Bourgoignie\inst{\ref{ROB}}
   \and
   L.~Dolla\inst{\ref{ROB}}
   \and
   C.~Jean\inst{\ref{ROB}}
   \and
   B.~Nicula\inst{\ref{ROB}}
   \and
   S.~Shestov\inst{\ref{ROB},\ref{CSL}}
   \and
   C.~Galy\inst{\ref{CSL}}
   \and
   R.~Rougeot\inst{\ref{ESTEC}}
   \and
   J.~Versluys\inst{\ref{ESTEC}}
   \and
   J.~Zender\inst{\ref{ESTEC}}
    \and
   P.~Lamy\inst{\ref{LATMOS}}
   \and
   S.~Fineschi\inst{\ref{OATo}}
    \and
   S.~Gun\'ar\inst{\ref{CAS}}
    \and
   B.~Inhester\inst{\ref{MPS}}
   \and
   M.~Mierla\inst{\ref{ROB},\ref{RomanianAcademy}}
   \and
   P.~Rudawy\inst{\ref{UWroclaw}}
    \and
   K.~Tsinganos\inst{\ref{UoA}}
   \and
   S.~Koutchmy\inst{\ref{IAP}}\thanks{Deceased}
   \and
   R.~Howard\inst{\ref{APL}}
   \and
   H.~Peter\inst{\ref{MPS}}
    \and
   S.~Viv\`es\inst{\ref{CEA}}
   \and
   L.~Abbo\inst{\ref{OATo}}
    \and
   C.~Aime\inst{\ref{UNice}}
   \and
   K.~Aleksiejuk\inst{\ref{CBK}}
   \and
   J.~Baran\inst{\ref{CBK}}
   \and
   U.~B\k{a}k-St\k{e}\'slicka\inst{\ref{UWroclaw}}
    \and
   A.~Bemporad\inst{\ref{OATo}}
   \and
   D.~Berghmans\inst{\ref{ROB}}
   \and
   D.~Be\k{s}liu-Ionescu\inst{\ref{AIRA}, \ref{RomanianAcademy}}
   \and
   S.~Buckley\inst{\ref{Onsemi}}
   \and
   O.~Buiu\inst{\ref{IMT}}
   \and
   G.~Capobianco\inst{\ref{OATo}}
   \and
   I.~Cimoch\inst{\ref{PCO}}
   \and
   E.~D’Huys\inst{\ref{ROB}}
   \and
   C.~E.~DeForest\inst{\ref{SwRI}}
   \and
   M.~Dzie\.zyc\inst{\ref{N7S}}
   \and
   K.~Fleury-Frenette\inst{\ref{CSL}}
   \and
   S.~E.~Gibson\inst{\ref{HAO}}
   \and
   S.~Giordano\inst{\ref{OATo}}
   \and
   L.~Golub\inst{\ref{Harvard}}
   \and
   K.~Grochowski\inst{\ref{N7S}}
   \and
   P.~Heinzel\inst{\ref{CAS},\ref{UWroclaw}}
   \and
   A.~Hermans\inst{\ref{CSL}}
   \and
   J.~Jacobs\inst{\ref{CSL}}
   \and
   S.~Jej\v{c}i\v{c}\inst{\ref{Lubjana1},\ref{Lubjana2},\ref{CAS}}
   \and
   N.~Kranitis\inst{\ref{UoA}}
   \and
   F.~Landini\inst{\ref{OATo}}
   \and
   D.~Loreggia\inst{\ref{OATo}}
   \and
   J.~Magdaleni\'c\inst{\ref{ROB}, \ref{KUL}}
   \and
   D.~Maia\inst{\ref{UPorto}}
   \and
   C.~Marqu\'e\inst{\ref{ROB}}
   \and
   R.~Melich\inst{\ref{TOPTEC}}
   \and
   M.~Morawski\inst{\ref{CBK}}
   \and
   M.~Mosdorf\inst{\ref{N7S}}
   \and
   V.~Noce\inst{\ref{OATo}}
   \and
   P.~Orlea\'nski\inst{\ref{CBK}}
   \and
   A.~Paschalis\inst{\ref{UoA}}
   \and
   R.~Pe\v{r}est\'y\inst{\ref{Serenum}}
   \and
   L.~Rodriguez\inst{\ref{ROB}}
   \and
   D.~B.~Seaton\inst{\ref{SwRI}}
   \and
   L.~Short\inst{\ref{ROB}}
   \and
   J.-F.~Simar\inst{\ref{CSL}}
   \and
   M.~St\k{e}\'slicki\inst{\ref{CBK}}
   \and
   R.~S\o rensen\inst{\ref{OIP}}
   \and
   G.~Terrasa\inst{\ref{CSL}}
   \and
   N.~Van~Vooren\inst{\ref{OIP}}
   \and
   F.~Verstringe\inst{\ref{ROB}}
   \and
   L.~Zangrilli\inst{\ref{OATo}}
     }

  \institute{
  Solar--Terrestrial Centre of Excellence --- SIDC, Royal Observatory of Belgium, 1180 Brussels, Belgium\\
  \email{Andrei.Zhukov@sidc.be}
  \label{ROB}
  \and 
  Centre Spatial de Li\`ege, Universit\'e de Li\`ege, Av. du Pr\'e-Aily B29, 4031 Angleur, Belgium
  \label{CSL}
  \and
  European Space Research and Technology Centre, European Space Agency, Noordwijk, Netherlands
  \label{ESTEC}
  \and
  Laboratoire Atmosph\'eres et Observations Spatiales, 11 Boulevard d’Alembert, 78280 Guyancourt, France
  \label{LATMOS}
 \and
  National Institute for Astrophysics, Astrophysical Observatory of Torino, Pino Torinese, Torino, Italy
  \label{OATo}
 \and
  Astronomical Institute of the Czech Academy of Sciences, 251 65 Ond\v{r}ejov, Czech Republic
  \label{CAS}
    \and
            Max Planck Institute for Solar System Research, Justus-von-Liebig-Weg 3, 37077 G\"ottingen, Germany\label{MPS}     
  \and
  Institute of Geodynamics of the Romanian Academy, 020032 Bucharest-37, Romania
  \label{RomanianAcademy}
 \and
  Astronomical Institute, University of Wroc\l{}aw, Kopernika 11, 51-622 Wroc\l{}aw, Poland
  \label{UWroclaw}
  \and
  University of Athens, Panepistimiopolis, 157 84 Zografos Athens, Greece
  \label{UoA}
  \and
  Institut d'Astrophysique de Paris, Paris, France
  \label{IAP}
  \and
  Applied Physics Laboratory, Johns Hopkins University, USA
  \label{APL}
  \and
  IRFM, CEA, F-13108 Saint Paul lez Durance, France
  \label{CEA}
  \and
  Universit\'e C\^{o}te d’Azur, Centre National de la Recherche Scientifique, Observatoire de la C\^{o}te d’Azur, UMR7293 Lagrange,
Parc Valrose, 06108, Nice, France
  \label{UNice}
  \and
  Centrum Bada\'n Kosmicznych Polskiej Akademii Nauk, Poland
  \label{CBK}
    \and
  Astronomical Institute of the Romanian Academy, Bucharest, Romania
  \label{AIRA}
\and
  Onsemi, Cork, Ireland
  \label{Onsemi}
  \and
  National Institute for Research \& Development in Microtechnologies, Bucharest, Romania
  \label{IMT}
  \and
  PCO S.A., Warsaw, Poland
  \label{PCO}
  \and
  Southwest Research Institute, Boulder, CO 80302, USA
  \label{SwRI}
  \and
  N7 Space Sp. z o. o., Warsaw, Poland
  \label{N7S}
  \and
  High Altitude Observatory, National Center for Atmospheric Research, Boulder, CO, USA
  \label{HAO}
  \and
  Harvard-Smithsonian Center for Astrophysics, USA
  \label{Harvard}
\and
  Faculty of Education, University of Ljubljana, 1000, Ljubljana, Slovenia
  \label{Lubjana1}
  \and
  Faculty of Mathematics and Physics, University of Ljubljana, 1000 Ljubljana, Slovenia
  \label{Lubjana2}
  \and
  Center for mathematical Plasma Astrophysics, KU Leuven, Leuven 3000, Belgium
  \label{KUL}
  \and
  Faculty of Sciences, Porto University, Porto, Portugal
  \label{UPorto}
  \and
  Institute of Plasma Physics ASCR v.v.i. (TOPTEC), Turnov, Czech Republic
  \label{TOPTEC}
   \and
  Serenum a.s., Praha, Czech Republic
  \label{Serenum}
   \and
  OIP Sensor Systems n.v., Oudenaarde, Belgium
  \label{OIP}
             }

   \date{}

  \abstract
  {
   We describe the scientific objectives and instrument design of the ASPIICS coronagraph launched aboard the \mbox{Proba-3} mission of the European Space Agency (ESA) on 5 December 2024. \mbox{Proba-3} consists of two spacecraft in a highly elliptical orbit around the Earth. One spacecraft carries the telescope, and the external occulter is mounted on the second spacecraft. The two spacecraft fly in a precise formation during 6 hours out of 19.63~hour orbit, together forming a giant solar coronagraph called ASPIICS (Association of Spacecraft for Polarimetric and Imaging Investigation of the Corona of the Sun). Very long distance between the external occulter and the telescope (around 144~m) represents an increase of two orders of magnitude compared to classical externally occulted solar coronagraphs. This allows us to observe the inner corona in eclipse-like conditions, i.e. close to the solar limb (down to 1.099~$R_\odot$) and with very low straylight. 
   ASPIICS provides a new perspective on the inner solar corona that will help solve several outstanding problems in solar physics, such as the origin of the slow solar wind and physical mechanism of coronal mass ejections.}


   \maketitle
\section{Introduction}
\label{S-intro}

The solar corona is the outer part of the solar atmosphere. It is very hot (1--2~MK) compared to the photospheric temperature of 5770~K. The corona is continuously expanding into interplanetary space in the form of the solar wind. Occasionally, eruptive energy release events in the solar atmosphere lead to coronal mass ejections (CMEs). The origin of the solar wind and the initiation of CMEs are fundamental scientific questions that are still not fully resolved. 

When arriving at Earth, solar wind streams and especially CMEs may disturb the geomagnetic field and produce a geomagnetic storm. CME-driven shocks also accelerate energetic particles, which represent a danger to astronauts traveling to the Moon and Mars \citep{Schwenn2006, Bothmer2007, Temmer2021}. These effects, collectively called ``space weather'', make studies of the solar corona important for practical applications as well. 

The solar corona has been observed for millennia during total solar eclipses. The near-coincidence of apparent sizes of the Sun and Moon makes a total eclipse possible. However, this also means that an eclipse is rather short: 7 minutes and 40 seconds at most, and typically 2--3~minutes. During a few minutes of totality, the corona rarely shows any dynamical evolution. The misalignment of the Moon's orbit around Earth with respect to the Earth's orbit around the Sun makes total eclipses rare: at most two per year, usually one per year, and none in some years. 

Modern remote-sensing observations of the corona are generally made using two techniques\footnote{We do not discuss here the radio imaging of the corona \citep{Gary2023}, which only has a low spatial resolution.}. The first technique is to observe the lower corona in extreme ultraviolet (EUV) or X-ray emission. In this case, the corona is observed not only above the limb but also against the solar disk. The field of view is typically limited to less than 1.5~$R_\odot$ (where $R_\odot$ is the solar radius and the distance is measured from the Sun’s center): 1.27~$R_\odot$ for the Atmospheric Imaging Assembly \citep[AIA, see ][]{Lemen2012} aboard the Solar Dynamics Observatory mission \citep[SDO, see ][]{Pesnell2012}, and 1.67~$R_\odot$ for the Solar Watcher using Active Pixel sensor \citep[SWAP, see ][]{Berghmans2006, Seaton2013} aboard the PROBA2 mission.

The second technique is to observe the high corona with externally occulted coronagraphs, such as LASCO C2 \citep[Large Angle Spectroscopic COronagraph, see][]{Brueckner1995} aboard SOHO \citep[SOlar and Heliospheric Observatory, see][]{Domingo1995}, or COR2 \citep{Howard2008} aboard STEREO \citep[Solar-TErrestrial RElations Observatory, see][]{Kaiser2008}, with the field of view typically starting around 2.5~$R_\odot$. Between the typical fields of view of EUV imagers and externally occulted coronagraphs, there is generally a gap where the observations are difficult\footnote{This gap partially overlaps with the middle corona defined by \citet{West2023}.}. Historically, this gap has been filled with ground-based coronagraphs, observing the emission line corona in visible or near-infrared \citep{Altrock2004, Tomczyk2008} or observing white light \citep{Fisher1981, deWijn2012, StCyr2015}, but these are limited to daytime good-weather viewing and in the case of white light, to observing in polarized brightness (since the background sky brightness is largely unpolarized).

One may try to fill in the gap by expanding fields of view of EUV imagers using occultations \citep{Slemzin2008, Rochus2020, Auchere2023} or off-points away from the Sun \citep{Tadikonda2019, Seaton2021}. However, solar EUV emission is proportional to the integral of the square of the coronal electron density ($n_{\rm e}^2$) along the line of sight. As the coronal density rapidly decreases with height, the signal-to-noise ratio quickly decreases, so the required exposure times become long (e. g. 1000~s). This may lead to the smearing of propagating structures, which complicates studies of dynamic phenomena \citep{Auchere2023}. 

The coronal white-light emission is proportional to the line-of-sight integral of $n_{\rm e}$. It therefore decreases with height slower than the EUV emission, so observing the corona in white light may allow faster imaging with at least comparable signal-to-noise ratio. Such observations can be made with internally occulted coronagraphs, which generally have a field of view starting at lower heights than externally occulted coronagraphs. For example, the field of view of the LASCO C1 \citep{Brueckner1995} and SECCHI COR1 \citep{Howard2008} coronagraphs start at 1.1~$R_\odot$ and 1.4~$R_\odot$ respectively. However, internally occulted coronagraphs allow light from the full solar disk to enter through their entrance apertures, which generally leads to high straylight levels on the detector. 

Straylight in coronagraphs is very difficult to suppress when corona below 2.5~$R_\odot$ is observed. High levels of straylight mean that the signal-to-noise ratio, the contrast of small-scale features, and the effective spatial resolution become low. Total solar eclipse observations do not suffer from straylight issues, but they are rare and last only for a short time. 

In this paper, we report the scientific objectives and instrument design of the ASPIICS coronagraph (Association of Spacecraft for Polarimetric and Imaging Investigation of the Corona of the Sun) that was launched aboard the \mbox{Proba-3} formation flying mission\footnote{The name of the mission, ``Proba'', means ``Try!'' in Latin. An alternative, older variant of spelling, ``PROBA'' (meaning ``PRoject for On-Board Autonomy'') was also used for the line of small technology demonstration ESA missions that includes \mbox{PROBA-1}, PROBA2, \mbox{PROBA-V}, and \mbox{Proba-3}. Both variants of the spelling have been used in the past for these missions. In this paper, we adopt the spelling \mbox{``Proba-3''}.} of the European Space Agency (ESA) on 5 December 2024. ASPIICS is an externally occulted coronagraph designed to image the solar corona between 1.099~$R_\odot$ and 3~$R_\odot$ in white light and in two narrow passbands, thus covering the gap between typical fields of view of EUV imagers and classical externally occulted coronagraphs. In Sect.~\ref{S-concept}, we describe the general principle of the ASPIICS observations and explain how the straylight is decreased due to the unique formation flying concept. Section~\ref{S-objectives} provides a description of the scientific objectives of ASPIICS. Sections~\ref{S-mission} and \ref{S-design} report the mission profile and instrument design, respectively. Section~\ref{S-performance} summarizes the instrument performances measured during on-ground calibration. Section~\ref{S-operations} describes the concept of operations. Section~\ref{S-summary} provides a short summary of the instrument and mission perspectives.

\section{Mission concept}
\label{S-concept}

\begin{figure}[!ht]
\centering
\includegraphics[width=0.5\textwidth]{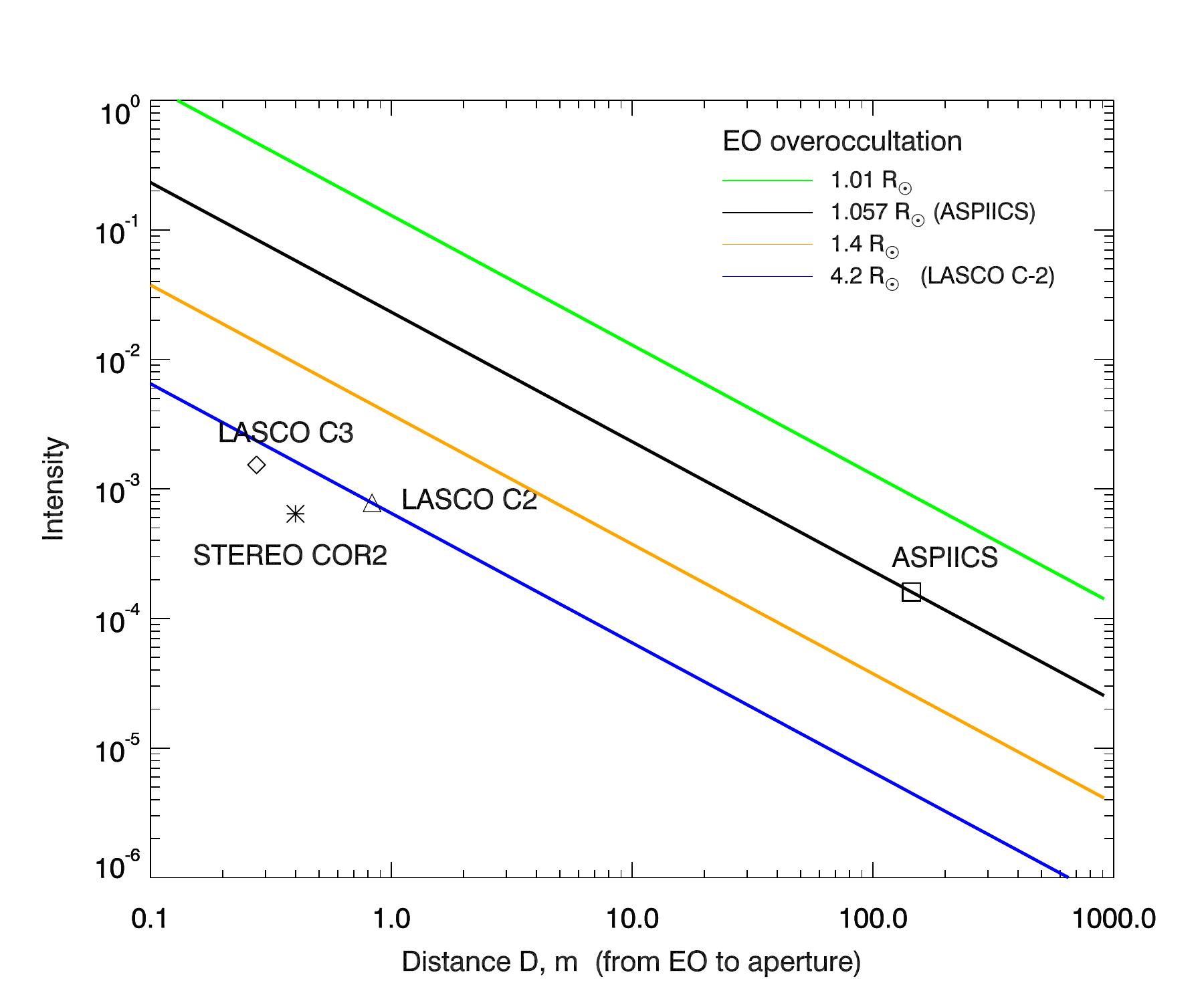}
\caption{Light diffracted on the external occulter (EO) of a coronagraph critically depends on the distance $D$ between the EO and the entrance aperture. The plot shows the dependence of the diffracted light intensity at the center of the aperture $L_{\rm A1}$ (normalized to the intensity of light that would be detected without the occulter) on $D$ according to Eq.~(\ref{Eq:diffraction})  for several angular sizes of the external occulter expressed in solar radii $R_{\odot}$, shown by slanted lines of different colors. In comparison with other externally occulted coronagraphs, the diffracted light intensity in ASPIICS is lower despite significantly smaller overoccultation.}
\label{fig:lenskii}
\end{figure}

In externally occulted coronagraphs, the solar disk light diffracted by the external occulter represents the major source of straylight \citep{Koutchmy1988, Aime2013, Rougeot2017, Rougeot2018, Shestov2018, Shestov2019, Shestov2021, DeForest2025}. According to theoretical calculations by \citet{Fort1978} and \citet{Lenskii1981}, the intensity $L_{\rm A1}$ (in relative units) of diffracted light in the center of the entrance aperture can be expressed as follows \citep[we use notations of ][]{Bout2000}:

\begin{equation}
    L_{\rm A1} = \Biggl\{\pi^2 r_{\odot} \Biggl[ 1 - \displaystyle{\Biggl(\frac{r_{\odot} D}{R_{\rm EO}}\Biggr)^2} \Biggr] \Biggr\}^{-1} \frac{\lambda}{R_{\rm EO}},
\label{Eq:diffraction}
\end{equation}
where $r_{\odot}$ is the angular radius of the Sun, $D$ is the distance between the external occulter and the entrance aperture, $R_{\rm EO}$ is the radius of the external occulter and $\lambda$ is the wavelength of observation. 

The intensity of the diffracted light for several representative situations is shown in Fig.~\ref{fig:lenskii}. The slanted lines show the diffracted light intensity for several angular sizes of the external occulter (expressed in $R_{\odot}$). For a given occultation, the diffracted light intensity decreases with increasing distance $D$ between the external occulter and the entrance aperture. The values for several classical externally occulted coronagraphs are shown in Fig.~\ref{fig:lenskii} as well\footnote{We note that Eq.~(\ref{Eq:diffraction}) is valid for a single occulter, and modern coronagraphs (LASCO C2 and C3, and COR2) have triple disk occulters. Eq.~(\ref{Eq:diffraction}) is used here to illustrate the physical principle of diffracted light calculation.}. The LASCO C2, C3, and STEREO COR2 coronagraphs\footnote{The Metis coronagraph \citep{Antonucci2020} has an inverted external occulter, which cannot be described by Eq.~(\ref{Eq:diffraction}).} are grouped around $D \sim 1$~m. For these missions, increasing $D$ beyond this value would be impractical due to considerations of the spacecraft size (of the order of a couple of meters for SOHO). This explains the use of internally occulted coronagraphs such as LASCO C1 and COR1 for observations of the inner corona from SOHO and STEREO, respectively. 

\begin{figure*}[!ht]
\sidecaption
\centering
\includegraphics[width=12cm]{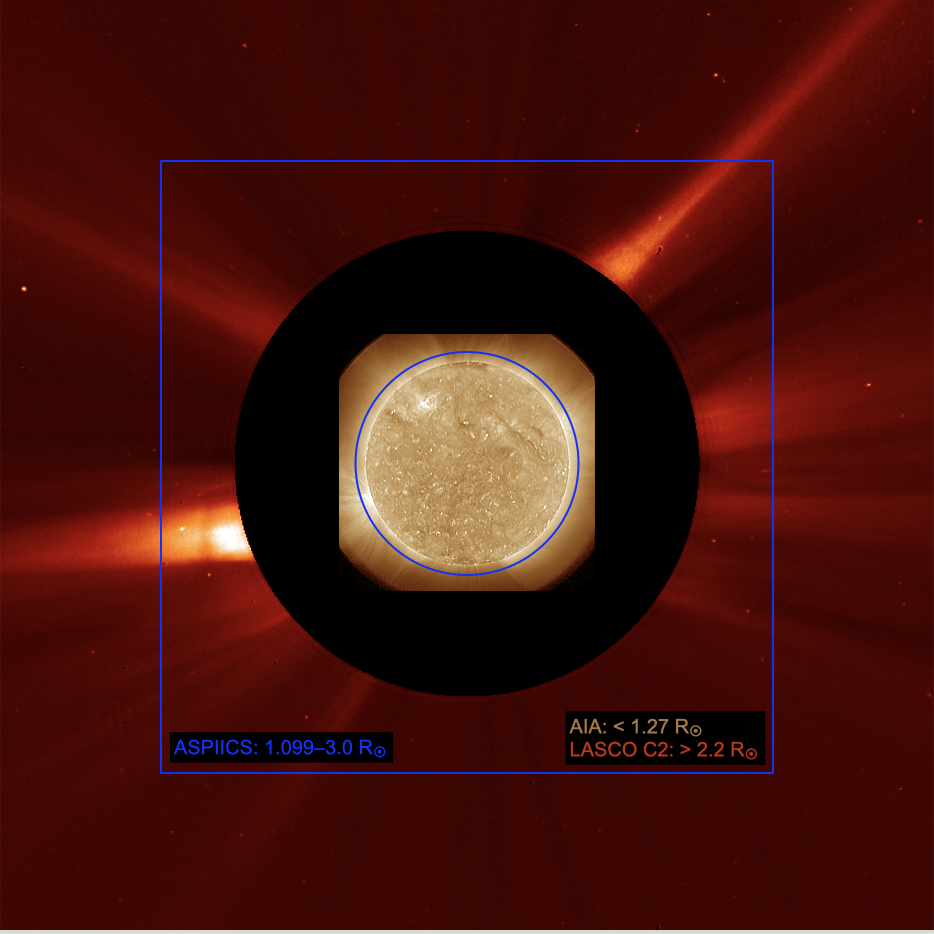}
\caption{Observations of the solar corona by classical externally occulted coronagraphs (SOHO/LASCO C2 image shown in red) and EUV imagers (SDO/AIA image shown in brown) leave an important  gap where observations are difficult. The field of view of the ASPIICS coronagraph aboard the \mbox{Proba-3} mission is shown in blue. }
\label{fig:fovs}
\end{figure*}

Due to the unique Proba-3 concept of two spacecraft flying in formation (see Sect.~\ref{S-mission}), the distance $D \approx 144$~m for ASPIICS, representing an increase by two orders of magnitude in comparison with classical coronagraphs. This allows us to achieve, in the same time, a better occultation (1.057 and 4.2~$ R_{\odot}$ for ASPIICS and LASCO C2, respectively\footnote{The occultation here represents the apparent size of the external occulter in solar radii, as seen from the entrance aperture. This is different from the position of the inner edge of the field of view, which also depends on the diameter of the entrance aperture due to vignetting \citep{Llebaria2004, Bayanna2011, Shestov2021}. The effect of vignetting may lead to the occultation being larger than the inner edge of the field of view, as it is e.g. for LASCO C2 with 4.2~$ R_{\odot}$ and 2.2~$ R_{\odot}$, respectively.}) and a more efficient rejection of diffracted light (by a factor 3--10). This demonstrates a unique advantage of ASPIICS. 

The main observations by ASPIICS are the coronal imaging in white light (wide spectral passband, 5363--5658~\AA ) between 1.099 and $3~R_{\odot}$ (see Fig.~\ref{fig:fovs}), thus allowing for a significant overlap with the low corona observations by EUV imagers \citep[e.g. SDO/AIA up to $1.27~R_{\odot}$, see ][]{Lemen2012}, and high corona observations by classical externally occulted coronagraphs \citep[e.g. SOHO/LASCO C2 from $2.2~R_{\odot}$, see ][]{Brueckner1995}. In addition to the total brightness images, polarized brightness images of the corona are taken with the same spectral filter. Polarized brightness data allow us to determine the coronal density more reliably than unpolarized data \citep[e.g.][]{Koutchmy1977, Decraemer2019, Lamy2020, Edwards2023}. Two filters with narrow spectral passbands are used, centered on the coronal green line (Fe XIV, 5304~\AA\ in vacuum) showing emission of the corona at temperatures around 2~MK \citep[e.g. ][]{Schwenn1997, Kim1997, Mierla2008, Ramesh2024}, and the He I D$_3$ line at 5877~\AA\ showing emission of prominences at temperatures around 0.01~MK \citep[e.g. ][]{Jejcic2018}. The images are projected on the square detector with 2048$\times$2048 pixels, with each pixel corresponding to 2.817\arcsec. The nominal synoptic cadence is 1~minute, although observations using a quarter of the field of view allow for higher cadences (up to 2~s).

\section{Scientific objectives}
\label{S-objectives}

The \mbox{Proba-3/ASPIICS} coronagraph examines the structure and dynamics of the corona in the crucial region between 1.1 and $3~R_{\odot}$, which is difficult to observe (Fig.~\ref{fig:fovs}). The top-level scientific objectives of ASPIICS are:

\begin{enumerate}
    \item Understanding the physical processes that govern the quiescent solar corona, by answering the following questions:

\begin{itemize}
    \item What is the nature of the solar corona on different scales?
    \item What processes contribute to the heating of the corona?
    \item What processes contribute to the solar wind acceleration?
\end{itemize}

\item Understanding the physical processes that lead to CMEs and determine space weather, by answering the following questions:
\begin{itemize}
    \item What is the nature of the structures that form the CME?
    \item How do CMEs erupt and accelerate in the low corona?
    \item What is the connection between CMEs and active processes close to the solar surface?
    \item Where and how can a CME drive a shock in the low corona?
\end{itemize}
\end{enumerate}

In the following sections, we describe each of the science objectives. 

\subsection{Understanding the physical processes that govern the quiescent solar corona}
\label{S-quiescent}

\subsubsection{What is the nature of the solar corona on different scales?}
\label{S-nature}

The large-scale structure of the solar corona is dominated by magnetically open dark coronal holes and bright streamers extending from closed coronal loops. During the years of low solar activity, a large coronal hole is situated around each pole, and helmet streamers constitute the streamer belt at low latitudes. During the epoch of high activity, polar coronal holes disappear and the corona is dominated by streamers all
around the solar disk \citep[e.g. ][]{Loucif1989}.

Due to the low-beta environment of the corona near the surface \citep{Gary2001, RodriguezGomez2024}, the plasma structure of the corona strongly depends on the configuration of the coronal magnetic field. Observations and models demonstrate that the magnetic configuration
of the streamer helmet may consist of one \citep{Saito1968, Koutchmy1971, Pneuman1971}, two \citep{Saito1973, Wang2007, Rachmeler2014} or three \citep{Schwenn1997, Noci1997, Banaszkiewicz1998, Wiegelmann2000} loop arcades. The arcades are situated below the bright quasi-radial streamer stalk, which represents the plasma sheet centered around the neutral current sheet
(for one or three arcades) or around a separatrix surface between magnetic fields of the same polarity (for two arcades -- so-called pseudostreamer). 

However, the coronal magnetic field is currently difficult to measure \citep[see e.g. reviews by ][]{DeglInnocenti2004, TrujilloBueno2010}. The global coronal magnetic
field is now routinely extrapolated from photospheric magnetograms
using models such as the potential field source surface (PFSS) model \citep{Schatten1969, Hoeksema1983}, and magnetohydrodynamic (MHD) models \citep{Linker1999, Yeates2008, Lionello2009, Downs2025}. The models do provide important information about the structure of the coronal magnetic field \citep[e.g. ][]{Wang2000GRL}: the positions of coronal holes (and of the heliospheric current sheet at 1~au) can be well derived from these models. However, determining the structure of streamers in this way may still be challenging \citep{Wang2002, Saez2005, Zhukov2008}. Even for the low solar activity epoch, the coronal field extrapolated from surface measurements is clearly model-dependent \citep{Yeates2010, Wiegelmann2017}. The extrapolated field is generally static and cannot account for slow evolution of streamer structure. Quite often, extrapolations fail to reproduce accurately complex topologies and dynamics inferred from remote-sensing observations of coronal structures, especially during high solar activity \citep[e.g. ][]{Zhukov2008, Benavitz2024}. 

The crucial transition between closed-field regions of the low corona (magnetic field dominated) and open-field regions higher up (solar wind dominated) is poorly observed by modern space missions  \citep[see e.g.][]{West2023}. 
Magnetic field topology can be inferred using observations of coronal structures \citep{Jones2016, Bemporad2023, Shi2024}. \mbox{Proba-3/ASPIICS} is well suited to provide crucial seamless observations of the low corona up to 3~$R_\odot$ in white light.  It can track the connectivity of coronal structures to the
solar surface and, in combination with state-of-the-art MHD models \citep[][]{Mikic2018, Downs2025}, allows us to determine reliably the coronal magnetic field configuration. 

On smaller scales, the solar atmosphere appears extremely structured and dynamic, in particular when observed at the best available spatial resolution. The small-scale structure and dynamics can be seen both in the lower corona \citep[e.g. ][]{Koutchmy1994, November1996, Delannee1998, Feldman1999, Zhukov2000, Cirtain2013, Druckmuller2014, Berghmans2021} and in the high corona \citep{Thernisien2006, DeForest2018, Decraemer2019, Poirier2020, Liewer2023}. The fine structure appears down to the limit of the spatial resolution of the instruments. Therefore, it is likely that the elementary structures have not yet been fully resolved. The spatiotemporal fine-structuring of plasma and magnetic field in the solar atmosphere determines the dissipation of energy and the fundamental physical processes leading to plasma heating, cooling, radiation, motion, and wave generation, and to solar wind and energetic particle acceleration. These processes, in turn, influence the fine structure of the corona. The dominant spatial and temporal scales of energy build-up, storage, and dissipation are presently
unknown.

ASPIICS can help answering the question about solar corona structuring and dynamics on different scales. Using its white-light and polarization brightness observations, ASPIICS is able to resolve density structures down to 5.63\arcsec\,  spatial resolution (4100 km, two pixels). This is largely superior to the resolution of other space coronagraphs like LASCO C2 (22.8\arcsec). Only Solar Orbiter/Metis can reach the same linear spatial resolution, but only during a short time close to Solar Orbiter perihelia \citep{Antonucci2020}. The high resolution of ASPIICS is available seamlessly from low to high solar corona (1.17--3~$R_\odot$), outside of the narrow vignetting zone (see Sect.~\ref{S-vignetting}).

Higher up, ASPIICS observations can be compared with the data from other coronagraphs (LASCO, Metis), and also with high-sensitivity images taken by the Wide-field Imager on Solar PRobe \citep[WISPR, ][]{Vourlidas2016}, which can reach heights down to 2.2~$R_\odot$ when the Parker Solar Probe mission \citep{Fox2016} is near its closest perihelia. Polarized ASPIICS data can be compared with observations taken by the Polarimeter to Unify the Corona and Heliosphere \citep[PUNCH, ][]{DeForest2026} in order to investigate the coronal density structures and their position in three dimensions \citep{Gibson2026}.

Sophisticated image processing algorithms \citep[e.g. ][]{Stenborg2003, Morgan2006, Druckmuller2014, Auchere2023WOW} will be used to improve the image quality further by increasing the contrast of coronal structures. Coronal density structure can be inferred using the polarized brightness inversions \citep[e.g. ][]{Lamy2020} and tomography \citep{AsensioRamos2023, Vasquez2024, WangT2025}. The magnetic configuration of the low corona can be inferred in its crucial region and compared with models \citep{Downs2025}. The excellent spatio-temporal resolution of ASPIICS allows us to investigate the fine structure and dynamics of the corona above the field of view of modern EUV imagers. 

\subsubsection{What processes contribute to the heating of the corona?}
\label{S-heating}

\begin{figure*}[!ht]
\sidecaption
\centering
\includegraphics[width=12cm]{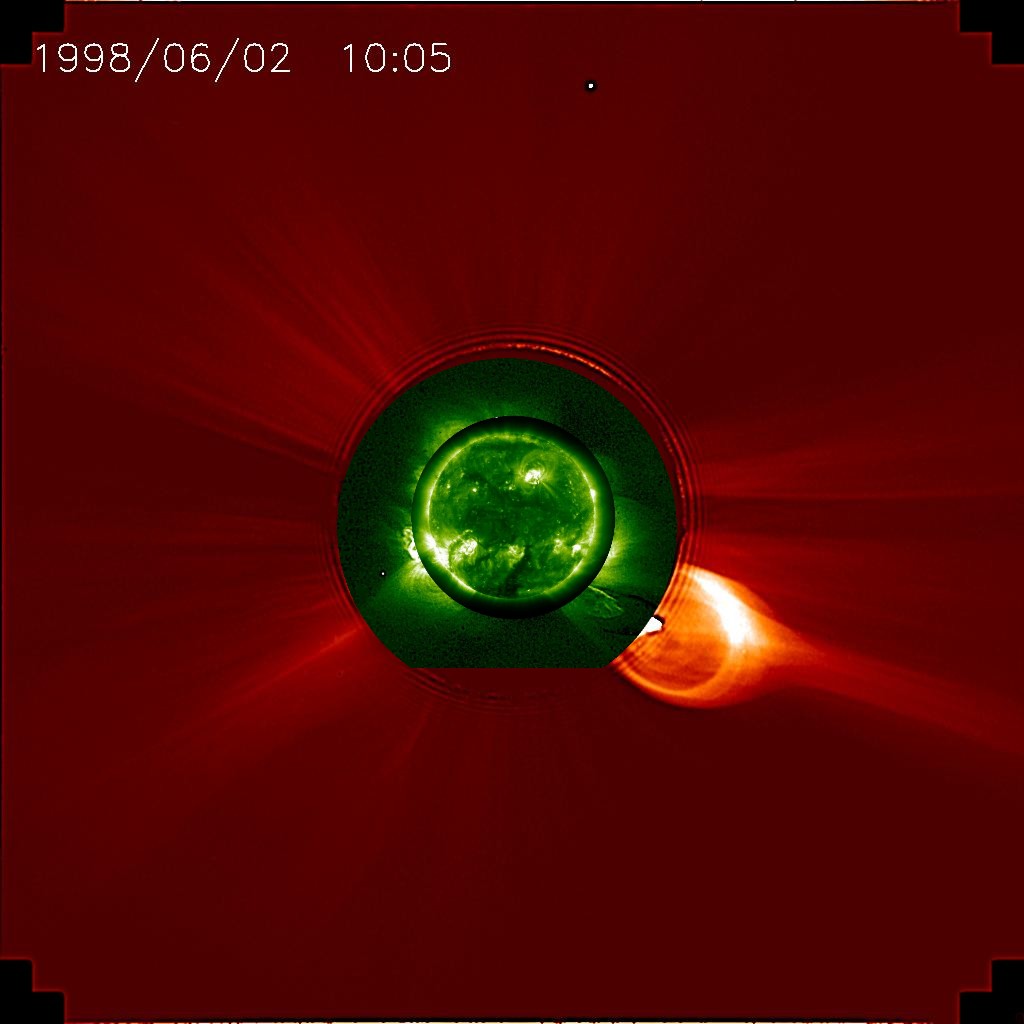}
\caption{Coronal mass ejection (CME) imaged by SOHO on 2 June 1998. The innermost
image is taken by EIT in the Fe~XII passband at 195~\AA, the middle image by LASCO C1 in
the Fe~XIV line at 5304~\AA, the outer image by LASCO C2 in white light. Contrast of small-scale
structures was increased using the algorithm by \citet{Stenborg2003}. Three-part CME
structure (evidence of a magnetic flux rope configuration) is well visible. ASPIICS can
image for the first time the CME dynamics seamlessly from 1.1 to 3.0~$R_\odot$, and nearly
simultaneously in white light, Fe~XIV and He~I D$_3$ passbands.}
\label{fig:cme}
\end{figure*}

The coronal heating problem is still one of the most debated questions in solar physics \citep[e.g. ][]{Parker1988, Klimchuk2006}. The consensus is that the coronal heating should be magnetic, i.e. it should occur by continuously extracting free magnetic energy from the dynamic corona.  Nanoflare theories explain coronal heating by dissipation of small-scale electric currents appearing due to photospheric motions \citep{Parker1972, Parker1988}. Wave theories consider dissipation of MHD waves launched by turbulent motions in the low atmosphere and propagating upwards \citep[e.g. ][]{Davila1987, VanDoorsselaere2007, Cranmer2005, Cranmer2007, vanBallegooijen2011}. ASPIICS, with its unprecedentedly high cadence (down to 2 s in white light), is very well suited to measure the wave properties in the corona and their contribution to the coronal heating.

In this section, we only consider the heating of the magnetically closed corona. Heating of the magnetically open corona needs to be considered together with the solar wind acceleration processes \citep[e.g.][]{Tu1997} and will be discussed in Section~\ref{S-wind}.

The photospheric magnetic field is highly structured, and each new generation of magnetographs resolves more numerous and smaller magnetic elements of mixed polarity \citep[e.g.][]{WangJ2022, Raouafi2023}. The photospheric field is dynamic, with photospheric magnetoconvection and magnetic field reconnection in the network being likely candidates to excite MHD waves \citep[e.g.][]{Axford1999, Nakariakov2020, WangT2021, Bale2023} that are truly pervasive in the solar atmosphere \citep[e.g.][]{Ofman1997, DeForest1998, Berghmans1999, Tomczyk2007, Andretta2025}. Due to a poor knowledge of local plasma and magnetic field parameters, it is difficult to determine if these
waves can be efficiently dissipated to heat the corona. Another important problem is how much of the wave energy flux generated in the lower atmosphere can propagate to the corona (and not e.g. be reflected at the transition region interface).

\begin{figure}[!ht]
\centering
\includegraphics[width=0.45\textwidth]{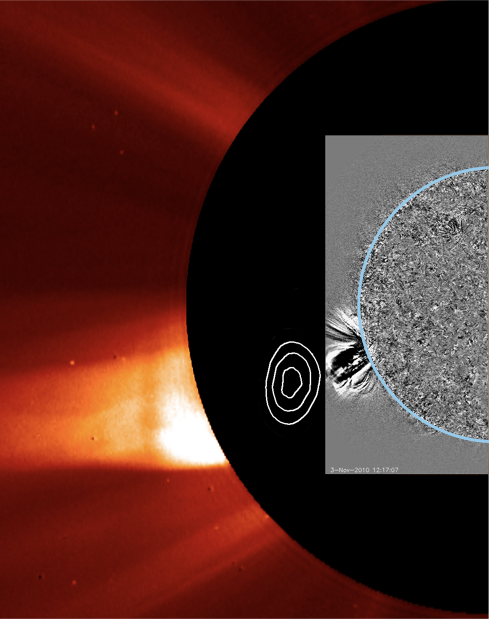}
\caption{Low-corona counterpart of a CME (bright front with a dimming inside) observed on 3 November 2010 by SDO/AIA in the 195~\AA\ passband (grey). This is a difference image, with a previous image subtracted to highlight the change of the coronal structure during the CME eruption. The blue circle shows the solar surface (1~$R_\odot$). Nearly simultaneous SOHO/LASCO C2 white light image is shown in red. Isocontours of the emission of a nearly simultaneous type II radio burst imaged by the Nan\c{c}ay Radioheliograph \citep{Kerdraon1997} are shown in the gap between the LASCO C2 and AIA fields of view. This illustrates the lack of coronagraphic observations in the gap region, which are provided by Proba-3/ASPIICS.}
\label{fig:shock}
\end{figure}

Recent observational efforts concentrated on the detection of Alfv\'en waves, which are theoretically considered to be the best able to transport the energy from the chromosphere to the corona. The wave flux in the chromosphere was estimated to be large enough to satisfy the coronal heating energy requirements \citep{DePontieu2007, Jess2009}. Estimates of the Alfv\'en wave flux in the corona are more
controversial, with perhaps a large energy flux that might be contained in still unresolved waves \citep{Tomczyk2007, McIntosh2011, Bailey2025, Morton2025}. Unlike linear Alfv\'en waves, which are non-compressive, non-linear Alfv\'en waves can drive density fluctuations \citep{Hollweg1971} that may be detected by ASPIICS due to its
unique capabilities (low straylight, large dynamic range and high signal-to-noise ratio). High spatial and temporal resolution of ASPIICS is perfectly suited to detect coronal waves via observations of quasi-periodic intensity perturbations \citep[e.g. ][]{Liu2011}. Complementary observations in the Fe~XIV passband and in white light mode may be very important as they provide constraints on thermal properties of oscillations and waves.

Except for the detection of the Kelvin-Helmholtz instability high in the corona \citep{Feng2013, Paouris2024} and CME-associated streamer waves \citep{Feng2011, Decraemer2020}, observations of coronal waves were up to now made only at rather small radial distances (up to around 1.25~$R_\odot$). ASPIICS can determine properties of coronal waves in its field of view up to 3~$R_\odot$. Combining electron densities derived from white-light ASPIICS measurements with temperature information obtained from the Fe~XIV passband data, one can estimate the wave flux as a function of height over an unprecedented radial range.

Advanced MHD models of the solar atmosphere \citep[e.g. Bifrost,][]{Gudiksen2011} include the upper convection zone, photosphere, chromosphere, and corona self-consistently. Convective motions below the surface lead to appearance of current sheets in the corona, which in turn leads to episodic and inhomogeneous heating via resistive dissipation. Both small-scale and large-scale heating can be modeled in this way, reproducing a number of properties of the hot corona, like temperatures and densities, and their evolution in response to different heating profiles \citep{Hansteen2015}. Current sheets can be generated through a turbulent cascade transporting energy to dissipation scales \citep{Rappazzo2008}. Subsurface and surface motions also lead to generation of Alfv\'enic fluctuations, which reach the dissipation scales via turbulent cascade \citep{Rappazzo2007, vanBallegooijen2011}. Depending on the loop parameters, different regimes of turbulence may develop and different scaling laws of the heating process may be observed \citep{Rappazzo2007, Rappazzo2008, vanBallegooijen2011, Knizhnik2018}. 

ASPIICS observations may be used to test the predictions made by these models. Information about distribution and evolution of coronal density can be obtained through observations of polarized brightness. Coronal temperature can be constrained using the observations in the green line. Temporal scales of coronal evolution and wave dynamics can be detected using high-cadence imaging. 

Due to novel observations of coronal waves and oscillations, ASPIICS can make an important contribution to the rapidly developing field of coronal seismology \citep{Nakariakov2005, Banerjee2007, Tomczyk2009, West2011, Feng2013}. Parameters
of waves and oscillations (e.g. frequency, amplitude, damping rate, the ratio of the first to the second harmonic) can be measured by ASPIICS with a very high precision, and at radial distances (between 1.099 and 3~$R_\odot$ solar radii) hardly accessible to current seismological studies \citep{Decraemer2020, Sorokina2024}. These observables can then be interpreted in terms of plasma and magnetic field characteristics. ASPIICS produces white-light images
that for the first time allow us to look at kink-mode oscillations of coronal structures (not only low coronal loops) from the perspective of density perturbations. ASPIICS seismological diagnostics complements its density diagnostics. ASPIICS can also image prominences in the He~I D$_3$ line, so
studies of waves inside of them and prominence seismology can be performed as well \citep[e.g. ][]{Okamoto2007}.

\subsubsection{What processes contribute to the solar wind acceleration?}
\label{S-wind}

\begin{figure*}[!ht]
\centering
\includegraphics[width=1.0\textwidth]{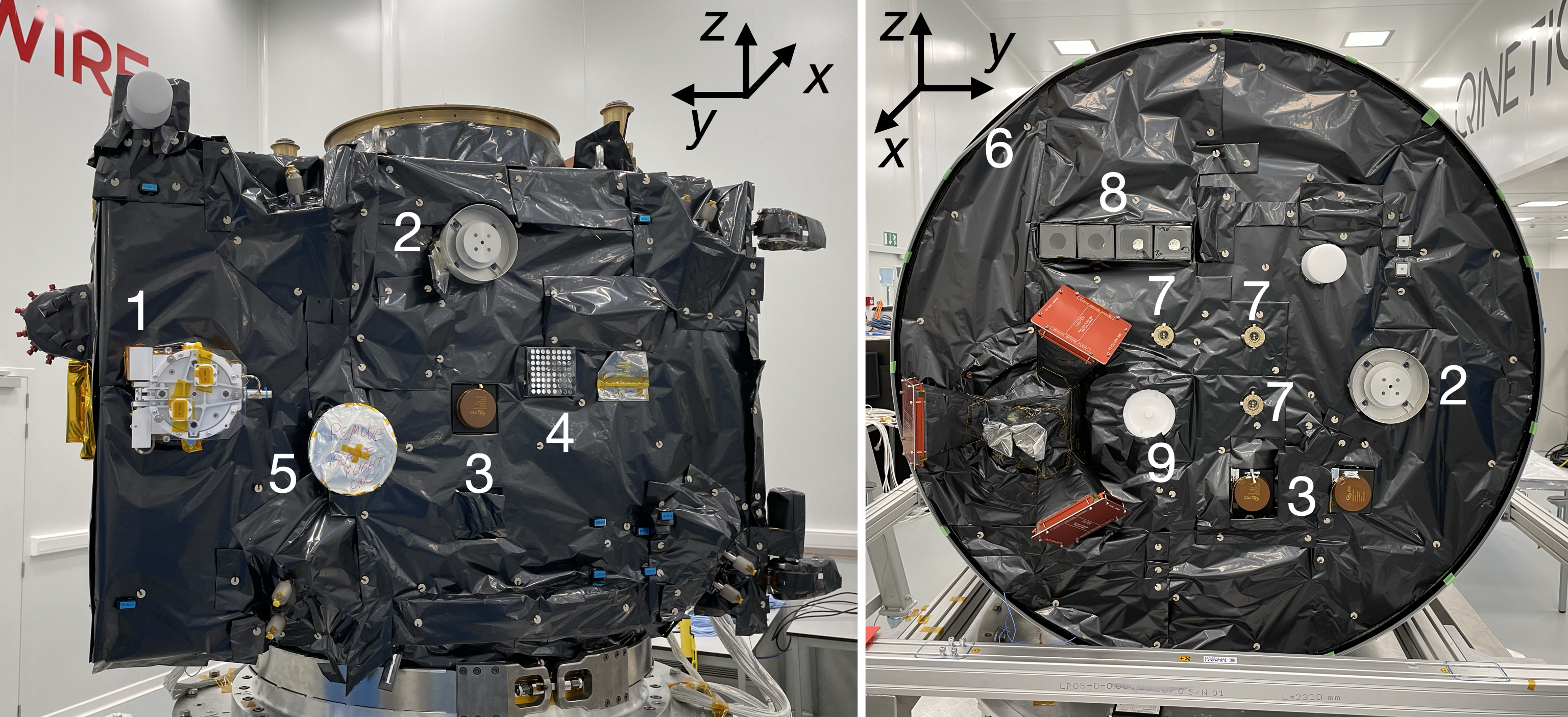}
\caption{Two spacecraft of the \mbox{Proba-3} mission. Left panel: the Coronagraph Spacecraft (CSC). Right panel: the Occulter Spacecraft (OSC). The annotations highlight key subsystems of the mission: the entrance door of the ASPIICS coronagraph (1), GNSS antennas (2), antennas of the Inter-Satellite Link (ISL, 3), some mires of the Visual-Based System (4), Corner Cube Retro-Reflector (5), which is a part of the Fine Lateral and Longitudinal Sensor (FLLS), the edge of the external occulter (6), three LEDs of the Occulter Position Sensor Emitter (OPSE, 7), wide-angle and narrow-angle cameras of the VBS (8), laser of the FLLS (9). The axes of the coordinate systems are shown in each panel, with the $x$-axis pointing away from the Sun, $z$-axis pointing towards the ecliptic north, and $y$-axis complementing the right-handed system. The theoretical formation corresponds to the perfect alignment of the respectively $x$, $y$, and $z$ axes attached to the two spacecraft.}
\label{fig:spacecraft}
\end{figure*}

Solar wind theory is well developed \citep{Parker2010, Lemaire2010}, and the overall structure of the heliosphere and its global solar boundary conditions are reasonably well understood from missions such as Helios, Ulysses and SOHO \citep[e.g. ][]{McComas2003}. During years of low solar activity, large polar coronal holes produce steady streams of fast solar wind with well-developed Alfv\'enic turbulence \citep{Bruno2013}, and their plasma originates in the expanding magnetic funnels in coronal holes \citep{Tu2005}. Slow wind flows originate from lower latitudes, in particular in the streamer belt \citep{Abbo2016}. During years of high solar activity, fast and slow wind can be observed at any latitude \citep{McComas2008}. Simple thermally-driven models are not able to explain fast solar wind speeds for reasonable values of the coronal temperature \citep{Parker1965, Hansteen2012, Halekas2022}, so additional acceleration mechanism is required.

The classical dichotomy between fast and slow solar wind is defined by wind speed measurements, which do not reveal the full picture. A more physical classification may distinguish a steady Alfv\'enic wind \citep{Belcher1971}, which may be fast or slow \citep{DAmicis2015} depending on the expansion factor of the magnetic field \citep[e.g. ][]{Wang2024, Ngampoopun2025}, and a variable slow wind with non-Alfv\'enic fluctuations \citep[e.g.][]{Belcher1971, Schwenn1990, Viall2015}. Alfv\'enic fluctuations are believed to originate via magnetic reconnection in the super-granulation network \citep{Axford1999}, or via photospheric motions \citep{Cranmer2005}. They propagate upward, with wave dissipation through a turbulent cascade contributing to plasma heating, and wave pressure providing additional solar wind acceleration \citep[e.g.][]{Cranmer2007}.

Another important consequence of photospheric magnetoconvection is the generation of small-scale jets due to interchange reconnection between open and closed field lines \citep{Raouafi2023}. Jets abundantly appear in the regions of mixed small-scale magnetic polarities that become visible in the magnetograms with the highest spatial resolution and sensitivity \citep{WangJ2022}. Small-scale jets were found to contribute significant amounts of mass and energy fluxes to the Alfv\'enic wind \citep{Raouafi2023, Chitta2023a}. Observations of EUV jets at the highest spatio-temporal resolution demonstrated their contributions to both fast and slow Alfv\'enic winds \citep{Chitta2025}. 

Recent observations by the Parker Solar Probe mission demonstrated that Alfv\'enic fluctuations in the near-Sun Alfv\'enic wind very often appear in the form of so-called ``switchbacks'', large-amplitude deviations of the magnetic field from the Parker spiral direction, which may even lead to the reversal of the radial field component \citep{Kasper2019, Bale2019, Horbury2020, Dudok2020}. Switchbacks appear in patches that can be realiably linked to the chromospheric network due to the close passage of Parker Solar Probe near the Sun \citep{Bale2021}. They seem to originate from small-scale interchange reconnection localized in the funnels of the super-granular magnetic network \citep{Bale2021, Bale2023}. These studies led to a picture of continuous interchange reconnection in the network being the major energy source for plasma heating and acceleration in the Alfv\'enic wind \citep{Bale2023}. This result was confirmed in a study of a radial alignment of Parker Solar Probe and Solar Orbiter spacecraft, which demonstrated that the energy lost by the Alfv\'en waves (mainly switchbacks) is consistent with the energy gained by plasma in solar wind heating and acceleration during its propagation between the corona and inner heliosphere \citep{Rivera2024}.

Helical morphologies similar to large-scale switchbacks were detected in the corona \citep{Telloni2022, Long2023}. Such large-scale switchbacks indicative of the interchange reconnection can be observed  by ASPIICS, which will provide observational constraints for their physical properties, in particular, spatio-temporal scales and densities, together with some indication on temperature from the Fe~XIV images. 

Solar wind acceleration mechanisms can also be probed by observing compressive fluctuations of the coronal brightness. The frequency spectra of intensity fluctuations are indicative of whether or not a turbulent cascade transfers energy to small scales where it can be dissipated to accelerate the solar wind. While magnetic and velocity field fluctuations can be reasonably described in terms of incompressible MHD turbulence, the nature of density fluctuations is still unclear \citep{Bruno2013}. The turbulence may already be developed in the corona and may carry the imprint of different wind sources.

The spectrum of density fluctuations can be obtained from coronagraphic measurements \citep{Bemporad2008, Telloni2009, Telloni2024} and compared to spectra obtained in situ. By applying a similar analysis to high-cadence (up to 2~s) time series of 
total and polarized white-light emission provided by ASPIICS, one can derive maps of density fluctuation
spectra of the nascent solar wind in the ASPIICS field of view below 3~$R_\odot$. Spectra of density fluctuations observed by ASPIICS can be compared to models \citep[e.g. ][]{Chandran2009} in order to understand the role of turbulence in the solar wind acceleration process.

The solar sources of non-Alfv\'enic, highly variable slow wind are the subject of ongoing debate \citep{Abbo2016, Cliver2025}. The slow wind speed in the corona was inferred from measurements of the speed of blobs in streamers \citep{Sheeley1997, Jones2009, Lyu2024, Alzate2024}. It is, however, unclear whether the variable slow wind originates from the interchange reconnection between open and closed field lines in streamers or pseudostreamers \citep{Wang2000, Higginson2017, Pellegrin2023, Romano2025}, from the
edges of streamers and near the streamer cusp \citep[e.g. ][]{Gosling1981, Wang1998, Reville2022}, from active region loop expansion \citep{Uchida1992}, or from open field near active regions \citep{Sakao2007, Slemzin2013}. Different mechanisms acting in different places at different times is also a possibility, which may explain high variability in the slow wind. 

Both coronal holes and streamer belt show a variety of small-scale dynamic events: detachments \citep{Koutchmy1973}, blobs \citep{Sheeley1997, Jones2009, Liewer2024, Liewer2025}, inflows \citep{Sheeley2007, Wang2018, Vourlidas2025}, jets \citep{Cirtain2007, Chitta2023b}, and other density structures \citep{Viall2015, DeForest2018}. Some of these phenomena are probably manifestations of interchange reconnection\footnote{This interchange reconnection occurs on a larger scale and higher in the corona than the interchange reconnection at the base of the Alfv\'enic wind described by \citet{Bale2023}.} between open and closed field lines in the solar wind source region \citep{Shibata1992, Pariat2009, Lynch2020}. The coronal hole boundary region is  expected to be a location of magnetic field disruption and reconnection, since many coronal holes rotate quasi-rigidly as compared to the differentially rotating photosphere \citep[e.g. ][]{Sime1986, Wang1996}. A narrow dynamic region around the open-closed field boundary can also be represented as a layer of separatrices and quasi-separatrices (so-called S-Web) that continuously releases solar wind due to interchange reconnection \citep{Antiochos2011, Baker2023, Wilkins2025}. 
Another important process is the flow-modified tearing mode instability at the heliospheric current sheet \citep{Reville2020}. It may lead to reconnection in the S-Web near the streamer cusp and the repeated ejection of small-scale flux ropes \citep{Reville2022}, which can be observed by coronagraphs \citep{Sheeley2009}. 
This continuous injection of mass and energy may be relevant to the solar wind acceleration \citep{Cirtain2007, Chitta2023a}, so understanding small-scale dynamics is crucial to solve the solar wind problem.

With ASPIICS, it is possible to make detailed investigations of small-scale dynamics in the solar wind source region, in particular at the interface between open and closed field domains (which is often also the interface between fast and slow wind). Due to its high spatial and temporal resolution in white light, ASPIICS can measure the fine structure and dynamics in unprecedented detail. ASPIICS can look for signatures of interchange reconnection and tearing mode instability in the heliospheric current sheet that can provide material for coronal streamers and slow solar wind. Due to its high spatial and temporal resolution, dynamic range and low straylight, ASPIICS can measure the slow solar wind speed continuously up to 3~$R_\odot$, thus providing a missing link between previous measurements \citep{Sheeley1997, Jones2009, Alzate2024} that are somewhat contradictory. This can help us to solve a long-standing problem of the origin of the variable slow solar wind. 

\subsection{Understanding the physical processes that lead to CMEs}
\label{S-CMEs}

CME initiation and evolution remain major themes in solar physics. Even though many CME initiation models coexist at present, the exact
processes that may cause a CME are still unknown. It is now clear that CMEs are related to a restructuring of the coronal magnetic field. The energy of a CME is stored in strongly non-potential field configurations
\citep[e.g. ][]{Titov1999} for a long time and then quickly released during an eruption due
to a loss of the magnetic field equilibrium \citep[][]{Low1996, Forbes2000}, or due to an instability \citep[e.g. ][]{Kliem2006}. 

\begin{figure}[!ht]
\centering
\includegraphics[width=0.45\textwidth]{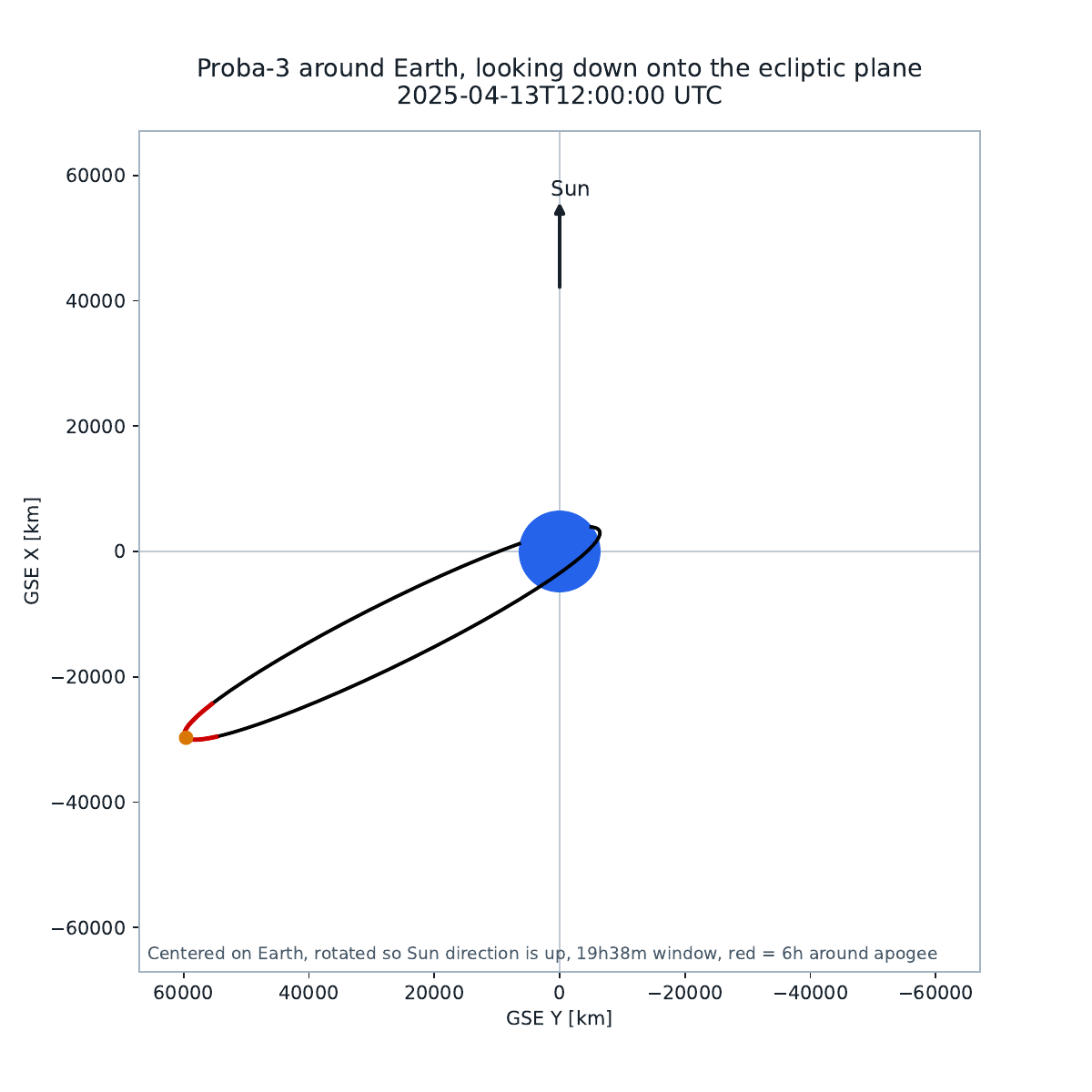}
\caption{The \mbox{Proba-3} orbit. The black line shows the 158th orbit of \mbox{Proba-3} around the Earth (blue circle) in the GSE (Geocentric Solar Ecliptic) coordinate system, seen in projection on the ecliptic plane from the positive direction of the GSE Z-axis. The part of the orbit corresponding to the 6-hour interval centered around the apogee passage, during which the formation flying is possible, is shown in red. The yellow dot shows the position of the \mbox{Proba-3} spacecraft on 13 April 2025 at 12:00~UTC. The direction towards the Sun is marked with a black arrow. The accompanying online movie shows the change of the orbit's orientation during the year (clockwise rotation in this view) due to Earth's revolution around the Sun. The orbital parameters are listed in Table~\ref{table:orbit}. }
\label{fig:orbit}
\end{figure}

\begin{figure}[!ht]
\centering
\includegraphics[width=0.45\textwidth]{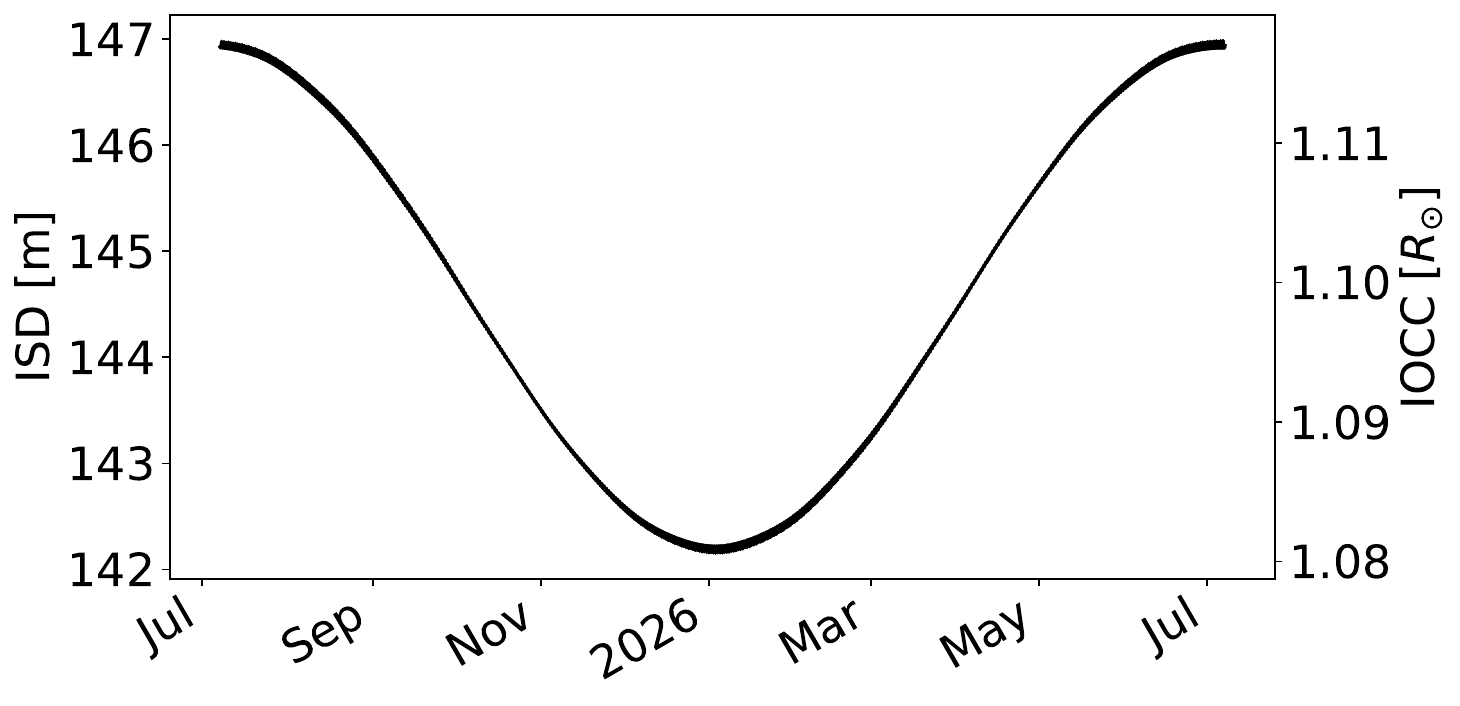}
\caption{Variations of the inter-satellite distance (ISD) over the year (left ordinate axis) and the limit of occultation by the internal occulter (position of the 100\% vignetting, right ordinate axis). The variations are nearly identical for both quantities, depending mostly on the angular size of the Sun that changes due to the eccentricity of the Earth's orbit. The occultation by the internal occulter fixed inside the telescope does not depend on the ISD.}
\label{fig:ISD}
\end{figure}

\subsubsection{What is the nature of the structures that form the CME?}
\label{S-structures}

\begin{figure*}[!ht]
\centering
\includegraphics[width=0.9\textwidth]{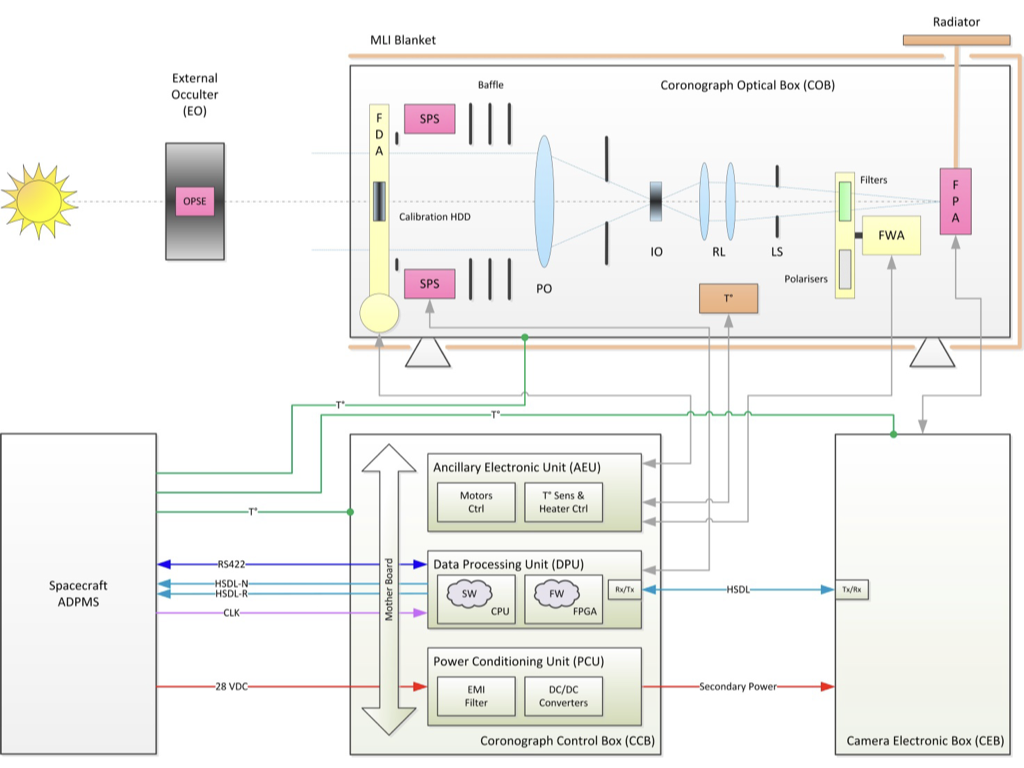}
\caption{The block diagram of the ASPIICS instrument. }
\label{fig:block}
\end{figure*}

In order to understand solar eruptions, knowledge of the state of the corona prior to a CME is essential, but the link of CMEs to underlying magnetic structures is not yet fully clear \citep[e.g.][]{Duan2024}. In white-light images, CMEs often show a three-part structure \citep{Illing1986}, composed of a bright leading edge
followed by a dark void and finally by a bright core associated with the erupting prominence (Fig.~\ref{fig:cme}). The three-part structure represents a two-dimensional projection of a three-dimensional magnetic flux rope \citep{Hundhausen1994, Low1995, Chen1997, Thernisien2009, Song2023}.

The simplest bipolar magnetic configuration of a pre-CME corona that can evolve into a magnetic flux
rope is that of a sheared arcade \citep{Priest1988, Cremades2004, Titov2008}. Alternatively, a pre-existing magnetic flux rope may also be present \citep{Canou2009, Cheng2013}. If the flux rope did not exist before the eruption, it is created by magnetic
reconnection once the CME starts \citep{Amari2011}. The flux rope ejection is accompanied by a large-scale magnetic field
restructuring \citep[e.g.][]{Roussev2007, Su2013}. More
complicated configurations of the pre-CME magnetic field are possible, e.g. two-arcade pseudo-streamers \citep{Torok2011, Wang2023, Wyper2024, Lynch2025} or three-arcade quadrupolar fields \citep{Antiochos1999, Lynch2008}. The basic physics of the loss of equilibrium or of an instability is believed to be the same regardless of the underlying magnetic configuration \citep[e.g. ][]{Wu2005, Kliem2006, Amari2007, Roussev2007}.

Prominence material can be aligned along the flux rope axis and thus serves as a marker for the
sheared arcade/flux rope configuration. It is isolated from the surrounding hot corona by the prominence
– corona transition region \citep[PCTR, ][]{Parenti2012, Gibson2018}. The study of prominences is an important topic in solar physics: their origin, equilibrium, thermodynamic properties, mass loading
and eruption are still not completely understood \citep[e.g. ][]{Heinzel2008, Berger2011}. Coronagraphic
observations of prominences are of great interest because the surrounding corona can be studied
simultaneously. ASPIICS can provide us with high-resolution, high-cadence simultaneous observations
of prominences in the He~I D$_3$ passband and of the surrounding corona in white light, complemented by the Fe~XIV images. The use of the He~I D$_3$ line in ASPIICS has the advantage that this line is optically thin. This permits a straightforward evaluation of the prominence
mass, as the emission in this line is well correlated with the emission in the H$_\alpha$ line. By accurately
evaluating the mass of erupting prominences, ASPIICS can provide us with crucial insight into the
problem of the CME mass budget \citep[e.g. ][]{Koutchmy2008}. Direct ASPIICS detection of the He~I D$_3$ emission in prominences can also help to infer the D$_3$ line polarization from Metis observations \citep{Heinzel2023}.

A cross-section of a sheared arcade can be observed in projection above the limb as a quiescent dark coronal cavity situated at the base of a helmet streamer \citep{Gibson2006, Jain2025}. Cavities are surrounded by bright streamer material and sometimes contain a prominence in their center \citep{Low1996, Marque2004}. As cavities show the locations of a simpler pre-eruptive magnetic configuration \citep[e.g.][]{Bak2013, Rachmeler2013}, understanding their structure and dynamics before and during eruptions is crucial for solving the CME initiation problem \citep{Gibson2015}.

Cavities are typically visible below 1.6~$R_\odot$, so their observations with space coronagraphs
are rare. Due to its seamless field of view from 1.099 to 3~$R_\odot$, high cadence and unprecedented
straylight rejection, ASPIICS is an ideal instrument to investigate coronal cavities and
their evolution into CMEs. Inversion of ASPIICS polarization brightness measurements can provide
densities inside and outside cavities \citep{Fuller2009}. The Fe~XIV images can be used to characterize the thermal state of the cavity plasma and ambient streamer material.
The He~I D$_3$ passband can be used to image the morphology and dynamics of the prominence
material often found inside the cavity \citep{Heinzel2008}. Properties of cavities, prominences inside of them and ambient streamer material derived by ASPIICS can be analyzed to understand their dynamics and relation to eruptions \citep{Song2025}, as well as the mass balance in the corona and in CMEs.

\subsubsection{How do CMEs erupt and accelerate in the low corona?}
\label{S-eruption}

\begin{figure*}[!ht]
\centering
\includegraphics[width=1.0\textwidth]{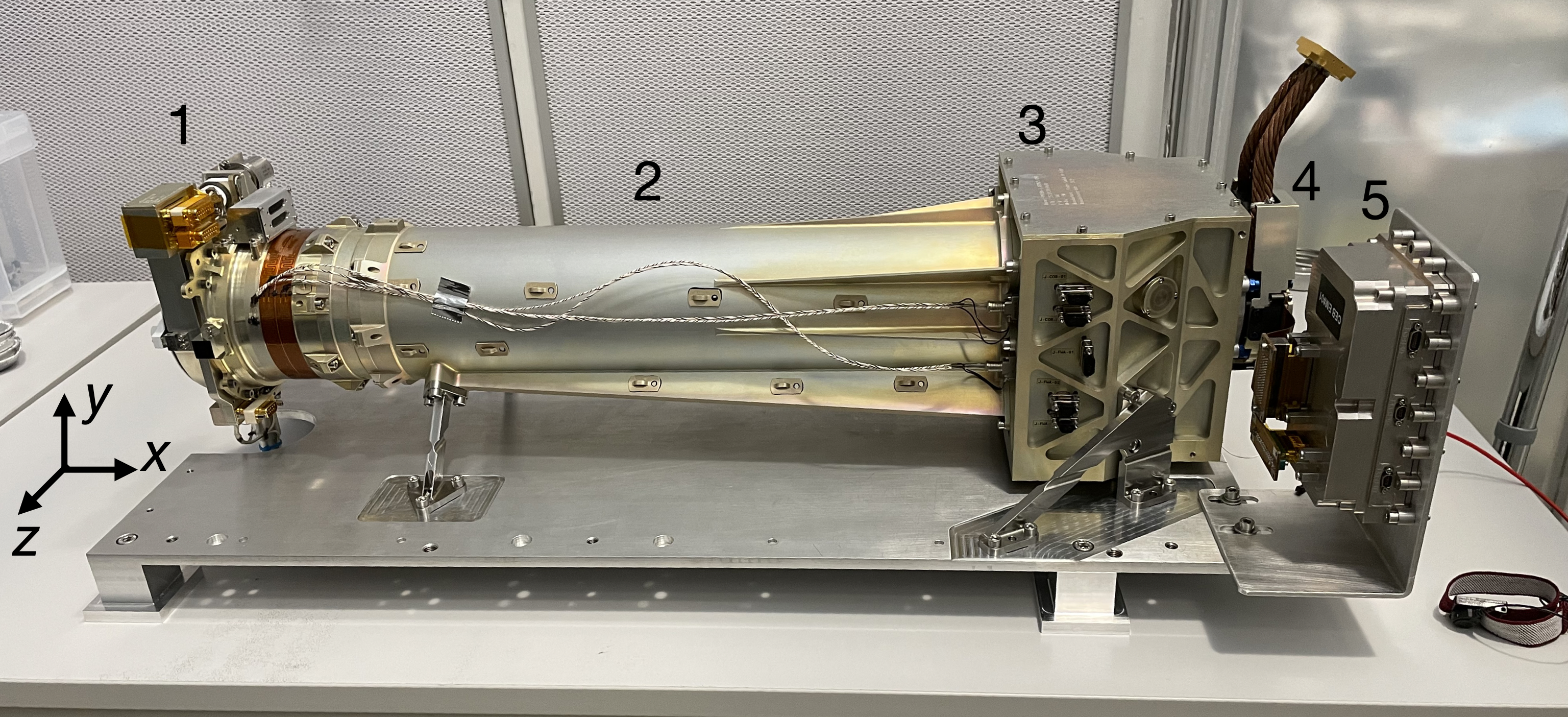}
\caption{A photo of the ASPIICS telescope. The annotations highlight key subsystems of the instrument: front door assembly (FDA, 1), telescope tube (2) carrying the lenses, equipment box (EQB, 3) that supports the filter wheel inside of it and the focal plane assembly (FPA, 4) outside of it, Camera Electronics Box (CEB, 5). The axes of the coordinate system attached to the instrument are shown, with the $x$-axis parallel to the optical axis of the telescope, $y$-axis directed from the bottom of the detector (row 0) to the top of the detector (row 2013), and the $z$-axis complementing the right-handed system. The ideal formation corresponds to the perfect alignment of the respectively $x$, $y$, and $z$ axes attached to the telescope and the two spacecraft (see Fig.~\ref{fig:spacecraft}).}
\label{fig:aspiics}
\end{figure*}

CMEs appear from magnetic field configurations that have enough of free magnetic energy accumulated during slow evolution of the coronal field in response to photospheric motions. The photospheric motions are too slow to drive most of the CMEs directly. The physics of the energy release process is still not entirely clear \citep[e.g.][]{Emslie2012}. At a certain moment in the evolution of the pre-eruptive magnetic field, an eruption may be triggered, e.g. due to such mechanisms as tether-cutting \citep{Moore2001}, magnetic breakout \citep{Antiochos1999}, kink instability \citep{Fan2003, Torok2005}, mass loading \citep{Wolfson1998, Seaton2011}. This brings the structure to an out-of-equilibrium state \citep{Kliem2014}. Once the force balance is lost, a driver mechanism leads to the fast acceleration and expansion of the CME. 

The driver of CMEs is the magnetic Lorentz force \citep[e.g.][]{Forbes2000, Zhong2023}. Two main types of driver mechanism are currently envisaged. The first is related to an ideal MHD instability \citep[e.g.][]{Kliem2014, Filippov2024}. Most probably it is the torus instability, which is due to the hoop force acting on a curved magnetic flux rope \citep{Kliem2006, Olmedo2010}. The torus instability leads to an eruption if the background magnetic field decreases with height quickly enough. The other possible driver mechanism is a resistive mechanism linked to magnetic reconnection under the erupting flux rope \citep[e.g. ][]{Lin2000, Karpen2012, Jiang2021}. The reconnection closes a part of the field overlying the flux rope. This disrupts the force balance and allows the eruption to continue, which leads to new magnetic flux being carried below the flux rope, where it reconnects again \citep{Lin2000}. This process thus involves a positive feedback between the flux rope eruption and reconnection underneath. It is still not clear which mechanism is dominant at which circumstances and at which stage of eruption \citep{Zhong2023, Xing2024}. It is also possible that different mechanisms are acting one after the other or simultaneously. Contributions of different triggering and driving mechanisms are difficult to disentangle. 

One of the problems in understanding the CME eruption and evolution in the low corona is that CMEs gain most of their acceleration below 3~$R_\odot$ \citep{MacQueen1983, StCyr1999, Vrsnak2001, Bein2011}, i.e. in a region not very well observed by past and present coronagraphs. CME kinematics is usually measured by three different types of instruments: EUV imagers like SDO/AIA at radial distances $r \lesssim 1.5~R_\odot$, internally occulted coronagraphs like STEREO/COR1 or SOHO/LASCO C1 at $1.2~R_\odot \lesssim r \lesssim 2.5~R_\odot$, and by externally occulted coronagraphs at $r \gtrsim 2.5~R_\odot$ \citep{Bein2011}. The low cadence of the LASCO C1 data allowed at best having only a handful of measurements during the CME impulsive acceleration \citep{Zhang2001}, and STEREO/COR1 data are subject to significant uncertainties due to high straylight \citep{Temmer2010}. These heterogeneous observations do not allow distinguishing between different proposed mechanisms of the CME initiation. Observations connecting ground-based with space-based coronagraphs have also demonstrated the importance of measuring acceleration from white-light observations of the low corona \citep{StCyr1999, Maricic2007, Bemporad2007, Bein2011}. Low straylight and a seamless view of the low corona up to 3~$R_\odot$ provided by ASPIICS can help to obtain a complete overview of the CME onset. ASPIICS high cadence and high spatial resolution are necessary for detailed measurements of the CMEs impulsive acceleration in the low corona, especially for
fast CMEs that today are not sufficiently well sampled in the critical range where the bulk of the acceleration occurs \citep{Veronig2018}. Speeds of different eruptive structures (e.g. erupting prominence, CME leading edge, multiple fronts) can be measured to determine any velocity dispersion in CMEs and its relation to the CME initiation mechanism \citep{Majumdar2024, Shaik2024, Bemporad2025}. 

ASPIICS can also contribute to solving the CME origin problem by observing in detail the restructuring of the large-scale magnetic configuration of the corona during CMEs. Its high cadence can allow precise measurements of timing of different events in its field of view: the start of the slow rise of the flux rope or prominence \citep{Cheng2020, Xing2024}, the onset of the fast reconnection in the flare current sheet below the erupting flux rope \citep{Jiang2021}, the start of the breakout reconnection above the flux rope \citep{Karpen2012}, and any additional reconnection of the erupting flux rope with other magnetic flux systems \citep{Lynch2025, Veronig2025}. This allows disentangling contributions of different physical mechanisms in  different phases of the eruption \citep{Xing2024}. 

\subsubsection{What is the connection between CMEs and active processes close to the solar surface?}
\label{S-connection}

\begin{figure}[!ht]
\centering
\includegraphics[width=0.45\textwidth]{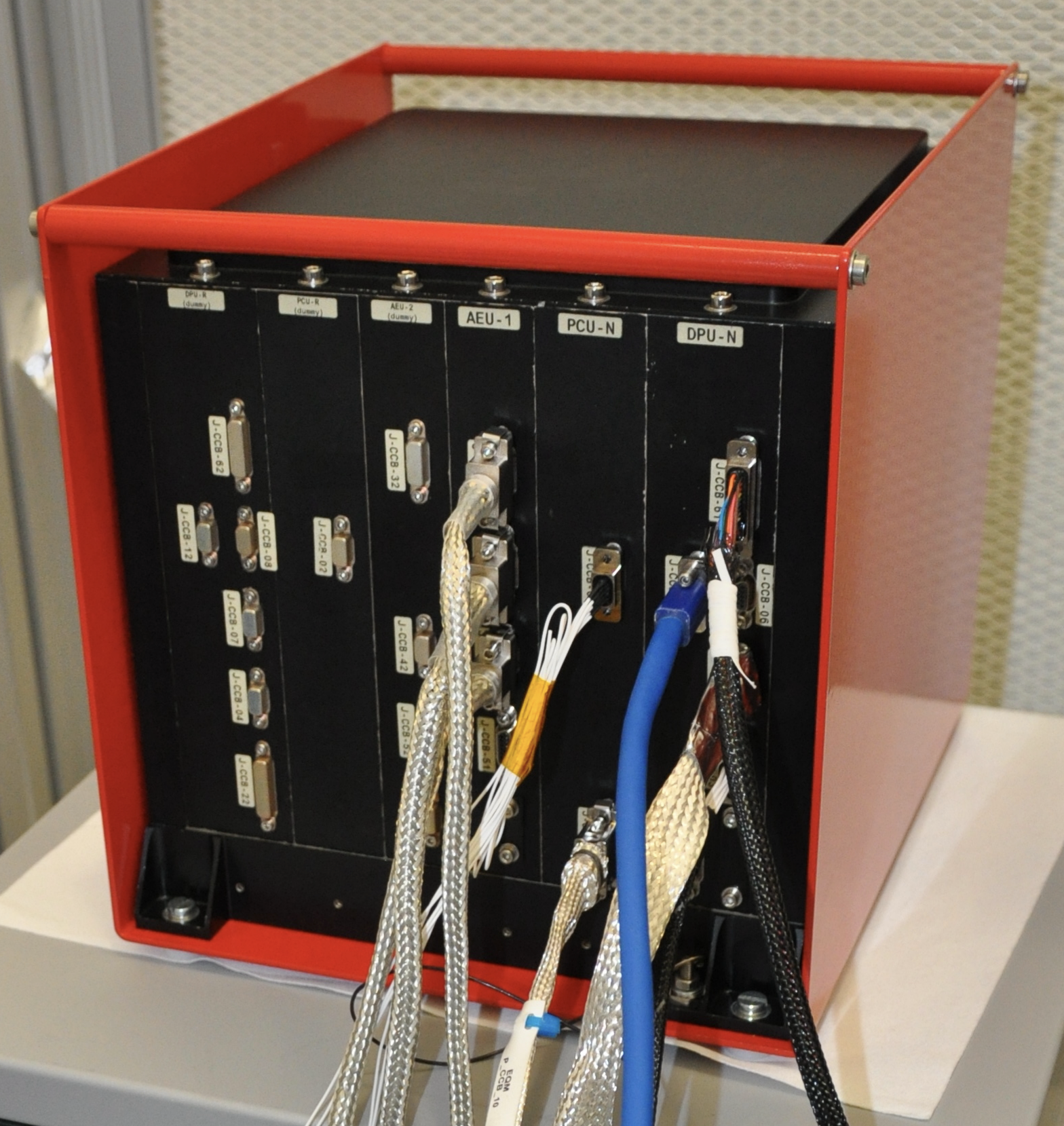}
\caption{A photo of the ASPIICS Coronagraph Control Box (CCB), the instrument computer.}
\label{fig:ccb}
\end{figure}

ASPIICS observations of the low corona can be an indispensable means to investigate the onset and early evolution of CMEs, and their association with other manifestations of solar activity. The link between CMEs and flares has been investigated since a long time \citep[e.g.][]{Gosling1993, Harrison1996}. It was demonstrated by \citet{Zhang2001} that the CME impulsive acceleration coincides with the rising phase of the associated soft X-ray flare. This correspondence
can be interpreted as evidence of a common cause of flares and CMEs – magnetic energy release \citep[e.g.][]{Emslie2012}. In particular, magnetic reconnection can diminish the magnetic tension of the arcade overlying an erupting flux rope and, at the same time, provide heating and accelerate energetic particles that produce observed flare signatures \citep[e.g.][]{Zhang2006, Temmer2010}. However, the synchronous evolution of the flare rising phase and CME acceleration is observed only in around half of CME-flare events \citep{Maricic2007}. High-cadence observations of ASPIICS can better constrain the CME acceleration curve and provide an important evidence on the association of CMEs with flares. 

As mentioned above, the pre-eruption configuration of CMEs is often associated with prominences. An eruptive prominence is often observed as the CME core \citep[e.g][]{Plunkett2000, Mierla2022}. Some prominence material in the CME core is seen to fall \citep{WangSheeley2002}. This mass loss can contribute to the acceleration of the CME in the framework of mass-loading CME models \citep[e.g. ][]{Wolfson1998, Seaton2011}. Prominences are also excellent markers of a slow rise of the pre-eruptive structure that can lead to a CME if it passes a certain height threshold \citep{Filippov2008}. ASPIICS measurements of prominence and CME mass in the crucial range of radial distances (up to 3~$R_\odot$) can be combined with measurements of eruption kinematics. Coronagraphic observations represent the only reliable way to measure the CME speed, so high-cadence ASPIICS observations can be
uniquely suited to determine the force balance in CMEs during their acceleration in the low corona. 

ASPIICS can provide a new view of such CME-associated dynamic
phenomena as coronal dimmings \citep[e.g.][]{Rust1976, Sterling1997, Zhukov2007, Veronig2025} and ``EIT waves'' \citep[e.g.][]{Thompson1998, Zhukov2011, Liu2014}. These phenomena are usually observed by EUV imagers. Due to complicated dependence of the EUV line intensity on temperature and density, plasma parameters in dimmings and EIT waves are difficult to evaluate \citep[e.g.][]{Zhukov2004}. The unique information on the
coronal density derived by ASPIICS can be combined with temperature information inferred from its Fe~XIV observations and from SDO/AIA for the coronal plasma diagnostics. Another crucial ASPIICS advantage
is its high cadence: radio observations demonstrate that plasma can
evolve very rapidly during a CME \citep[e.g.][]{Maia1999}. ASPIICS data may be thus crucial for our understanding of plasma
dynamics during the CME onset. 

Spectroscopic observations of CMEs by SOHO/UVCS showed the existence of persistent (up to a few days) high-temperature (more than 8 MK) emission in the CME wake \citep{Ciaravella2002, Raymond2003, Bemporad2006}. This unusually hot plasma was ascribed to the heating by reconnection in the post-CME current sheet predicted by CME models \citep[see the review by ][]{Forbes2000}. In white light images, current sheets generally correspond to bright ray-like structures \citep[see e.g.][]{Ko2003, Webb2003, Ding2024}, often exhibiting dynamic fine structure, like blobs and plasmoids \citep{Patel2020}.

However, this interpretation leaves several open problems that can be directly addressed by ASPIICS. The thickness of the reconnection layer is expected to be comparable to the ion Larmor radius (of the order of
a few meters in the corona). In contrast, the thickness of post-CME radial structures observed both in white-light and EUV emissions is typically of the order of 10$^4$--10$^5$ km, and the projection effects are too weak to account for this discrepancy. Another problem is the unexpectedly long duration of post-CME current sheets.
Both problems may point at an effective resistivity much larger than even the anomalous resistivity. It can be achieved in the turbulent current sheet subject to e.g. the tearing mode instability \citep{Lin2007, vanBallegooijen2008}. Important evidence in favor of this idea was obtained in UV observations \citep{BemporadA2008, Susino2013}, although the hard X-ray data indicate the possibility of plasma supply from a hot long-lived source \citep{SaintHilaire2009}. Another possibility is an effectively ``ideal'' tearing mode appearing at very high Lundquist numbers \citep{Pucci2014, Tenerani2015}, which may lead to fast reconnection even without anomalously high resistivity.

ASPIICS can provide unique observations of post-CME current sheets due to its combination of a large field of view (including
the whole region where these current sheets form and evolve), high spatial resolution, and cadence. ASPIICS allows us to measure the current sheet plasma densities (polarization brightness measurements) and to derive at the same time information on temperature (from Fe~XIV images) at different altitudes along the current sheet during its whole lifetime. A deeper knowledge on post-CME current sheets provided by ASPIICS can be of significant importance for our understanding of the fundamental process of magnetic reconnection in astrophysical conditions.

\subsubsection{Where and how can a CME drive a shock in the low corona?}
\label{S-shocks}

Generation of large-scale shock waves in the solar corona, their propagation to the interplanetary space and possibility of energetic particle acceleration by them are major questions in the science of solar-terrestrial relations \citep[space weather, see e.g.][]{Schwenn2006}. Two physical explanations are proposed for the generation of a coronal shock: a flare blast wave and a piston-driven shock due to a CME \citep{Vrsnak2008}. It is generally accepted that non-corotating interplanetary shock waves are CME-driven \citep[e.g.][]{Schwenn2006}, but the origin of coronal shocks is still a matter of debate. Both CME and flare origins seem to be possible \citep[e.g.][]{Wagner1983, Cliver1999, Gopalswamy2009, Magdalenic2010, Magdalenic2012, Jarry2023}.

Observing shock waves by a coronagraph is a complex task due to often-insufficient sensitivity of the instrument and to the difficulty in associating features observed in white light with shocks. White-light shocks are believed to appear as weak bright fronts, typically ahead of some fast CMEs \citep{MacQueen1983, Sheeley2000, Eselevich2008, Ontiveros2009, Lu2017} . Another important shock signature
is streamer deflection \citep{Sime1987, Sheeley2000, Decraemer2020}. In both cases, however, it is not clear if the wave is a true shock as plasma and magnetic field parameters in
the CME and in the ambient corona are poorly known \citep[e.g.][]{Frassati2024}. \citet{Vourlidas2003} used an MHD simulation
to confirm that the density enhancement in front of a CME indeed represents a fast-mode shock.
They, however, could not observe it below 2.2~$R_\odot$, so it was impossible to determine the moment of the shock formation. Apart from visual inspection of white-light images, CME-driven shocks
can be identified in the SOHO/UVCS data with the appearance of broad profiles of the O~VI line \citep{Ciaravella2005}. Halo CME fronts were found to correspond to coronal plasma swept up by a
shock or a compression wave \citep{Ciaravella2006, Vourlidas2013}.

Shocks in the low corona can also be detected
by observing type II radio bursts in solar
dynamic radio spectra \citep[e.g.][]{Cliver1999}. Radio
imaging demonstrates that type II bursts first appear
very low in the corona (1.2--1.5~$R_\odot$), see e.g.
\citet{Vrsnak2005, Carley2013}. In addition
to the problem of choosing a coronal density model to
estimate the shock speed from the drift rate of type II
bursts, the lack of coronagraphic CME observations in
this range of radial distances (see Fig.~\ref{fig:shock}) strongly impedes our
efforts to solve the problem of the origin of coronal shocks \citep{Magdalenic2010}. STEREO/COR1 observations
between 1.1 and 4~$R_\odot$ provided us
with an important insight \citep{Gopalswamy2009},
but the lack of concurrent radio imaging observations
from the same vantage point is a strong limitation of
such studies.

ASPIICS can provide us with important evidence
on the origin of coronal shocks due to its unprecedented observations of the region below 3~$R_\odot$, where many shocks originate. High cadence
white-light ASPIICS observations, in combination
with high-cadence ground-based radio spectra and
images of type II bursts, may allow us to understand the shock formation in the low corona. Polarization
brightness and white-light observations by ASPIICS can also be used to derive better density
models that are crucial to determine the shock speed from type II burst observations in dynamic spectra.

\section{Mission profile}
\label{S-mission}

The first attempt to observe solar corona from space using formation flying was made in 1975 during the Apollo-Soyuz Test Project mission by the United States and Soviet Union. A camera mounted aboard the Soyuz spacecraft was used to take photos of the corona, while the solar disk was occulted by the Apollo spacecraft. The F-corona was detected \citep{Nikolskiy1977}. 

\mbox{Proba-3} was launched on 5 December 2024 aboard the PSLV-XL rocket (Polar Satellite Launch Vehicle) from Satish Dhawan Space Centre, the main spaceport of the Indian Space Research Organization (ISRO) located in Sriharikota, India. The mission duration is planned to be two years, including four months of the commissioning. \mbox{Proba-3} consists of two spacecraft (Fig.~\ref{fig:spacecraft}): the coronagraph spacecraft (CSC) and the occulter spacecraft (OSC), see e.g. \citet{Llorente2013} and \citet{Peters2014}. The CSC carries the ASPIICS telescope and the OSC carries the circular external occulter disk with a toroid edge. The spacecraft are placed in a highly elliptical orbit (HEO\footnote{This acronym can be also expanded to Highly Eccentric Orbit, which is perhaps more precise.}) around the Earth (Fig.~\ref{fig:orbit}). The basic parameters of the orbit are listed in Table~\ref{table:orbit}. The orbital period is 19 hours 38 minutes, and during 6 hours centered around the apogee, the two spacecraft can perform precise formation flying, allowing the entrance aperture of ASPIICS to be placed in the shadow created by the external occulter. The inter-satellite distance (ISD) is around 144~m, varying with the change of the Sun-Earth distance during a year as follows:

\begin{equation}
    D = \frac{R_{\rm EO} - R_{\rm A1}}{\tan{(C r_{\odot})}} - \frac{T}{2} - R_{\rm c}\sin{r_{\odot}},
\label{Eq:ISD}
\end{equation}
where $R_{\rm EO} = 0.71088$~m is the radius of the external occulter, $R_{\rm A1} = 0.025$~m is the radius of the entrance aperture, $C = 1.02$ is the occultation factor by the external occulter (edge of the 100\% vignetting), $r_{\odot}$ is the apparent radius of the Sun, $T = 0.035$~m is the thickness of the external occulter disk, $R_{\rm c} = 0.7$~m is the radius of curvature of the toroid edge of the occulter disk. Figure~\ref{fig:ISD} shows the change of the ISD during the year, which is primarily produced by the change of the apparent radius of the Sun $r_{\odot}$ due to eccentricity of the Earth's orbit. 

The precision of the alignment of the spacecraft in the longitudinal and lateral directions is around 10~mm and 1~mm, respectively. The radius of the geometric umbra in this configuration is 39.6~mm. 

\begin{table}[!ht]
\renewcommand{\arraystretch}{1.2}
\begin{center}
\caption{Parameters of the \mbox{Proba-3} orbit at the beginning of the mission}
\label{table:orbit}
\begin{tabular}{cc}    
  \hline\hline                   
Parameter name & Parameter value \\
  \hline
Perigee height & 600 km \\
Apogee height & 60530 km \\
Semi-major axis & 36942.96 km \\
Eccentricity & 0.8111 \\
Inclination & 59$^{\circ}$ \\
 RAAN\tablefootmark{a} & 142$^{\circ}$--148.1$^{\circ}$ \\
Argument of perigee & 188$^{\circ}$ \\
Orbital period & 19 h 38 min \\
  \hline
  \tablefoottext{a}{Right ascension of the ascending node.}
\end{tabular}
\end{center}
\end{table}

\begin{figure*}[!ht]
\centering
\includegraphics[width=1.0\textwidth]{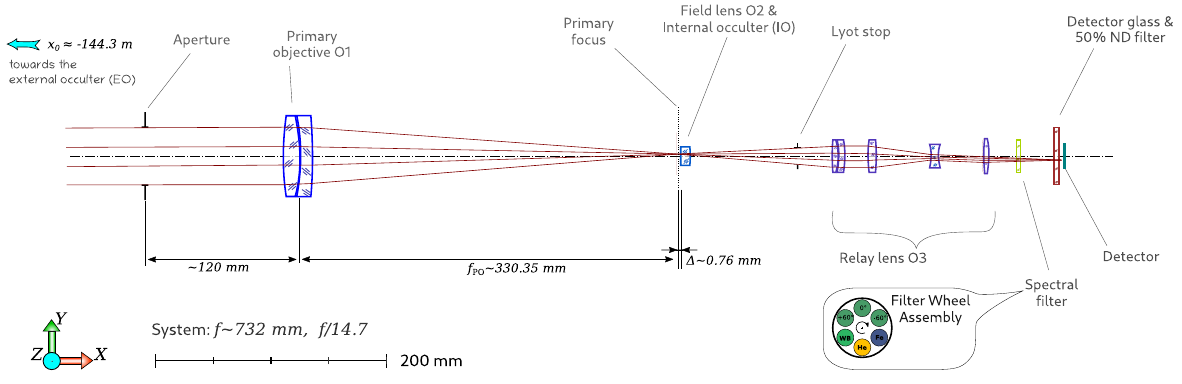}
\caption{Optical scheme of the ASPIICS instrument. }
\label{fig:optics}
\end{figure*}

To maintain the two spacecraft in formation, a number of technologies are used. First of all, the positions and velocities of the two spacecraft are refined using relative GNSS (Global Navigation Satellite System) when the spacecraft pass near the orbital perigee \citep{Ardaens2013}. After the GNSS visibility is lost, this information is propagated further in time, up to the availability of the data from the Visual-Based System (VBS) mounted on the OSC. The VBS consists of a Wide Angle Camera (WAC) and a Narrow Angle Camera (NAC), which are used to detect nine blinking mires (LEDs) on the CSC and thus image the CSC directly during flight. The Fine Lateral and Longitudinal Sensor \citep[FLLS, see][]{Bradshaw2018} includes a beam produced by an infrared laser (working wavelength 980~nm) mounted on the OSC. The laser beam is reflected by the CCRR (Corner Cube Retro-Reflector) placed on the sun-facing side of the CSC. The reflected laser beam is captured by the optical head unit of the FLLS for precise lateral and longitudinal measurement of the CSC position with respect to the OSC. Adjustment of position is performed, if necessary, using thrusters. The Inter-Satellite Link (ISL) is a radio communication means used by the two spacecraft to exchange information and coordinate their maneuvers. 

The formation flying can be maintained autonomously over 6~hours in every orbit. This corresponds to around a factor 100 improvement in the duration of uninterrupted observations in comparison with a total eclipse observed on the ground (maximum 7~minutes 40~seconds). The baseline total observation time is determined by the available fuel and is currently estimated to constitute 1000~hours over the 20-month duration of the nominal mission. On average over the mission duration, \mbox{Proba-3} will observe the corona more than two orbits per week, which constitutes around a factor of 50 improvement in the occurrence rate in comparison with a total eclipse (at most once per year, rarely twice a year).

Five ground stations (in Spain, Chile, Australia, and French Guyana) are used to downlink the science, housekeeping, and ancillary data. The data downlink rate is highly irregular, as it varies with such factors as the visibility of the spacecraft from a particular ground station, the ground station availability for Proba-3, and the distance to the ground station (which varies due to a highly elliptical orbit). The maximum science data volume is around 30 Gb per orbit.

\section{Instrument design}
\label{S-design}

Early works on instrument design were reported by \citet{Lamy2007}, \citet{Lamy2010}, and \citet{Vives2010}, and the evolution towards the final design was described by \citet{Renotte2014} and \citet{Galano2018}. The ASPIICS block diagram is shown in Fig.~\ref{fig:block}, and a photo of the flight model of ASPIICS is shown in Fig.~\ref{fig:aspiics}. The external occulter mounted on the OSC provides the shadow for the entrance aperture of the telescope placed on the CSC. The entrance aperture is protected by a reclosable door (Front Door Assembly, FDA) that is open during observations. The instrument optics project the image of the corona on the CMOS (Complementary metal-oxide-semiconductor) Active Pixel Sensor (APS) detector that has 2048$\times$2048 pixels of 10~$\mu$m size \citep{Renotte2016}. The temperature of the detector is controlled by the radiator that evacuates the excessive heat to an open space. The detector output is read by the Camera Electronic Box (CEB) that transfers the data further to the Coronagraph Control Box (CCB, see Fig.~\ref{fig:ccb}). The CCB provides the power to the coronagraph and controls the CEB, the FDA, and the Filter Wheel Assembly (FWA). It also sends the data to the spacecraft computer (Advanced Data and Power Management System, ADPMS), from where they are downlinked to the ground. 

\subsection{Optical design}
\label{S-optics}

ASPIICS is a classic externally occulted Lyot coronagraph. The external occulter is a disk made of CFRP (Carbon Fiber Reinforced Polymer) with the radius $R_{\rm EO} = 0.71088$~m mounted on the OSC spacecraft. The disk has a toroid edge, which keeps the apparent occulting disk radius constant in the event of a small (up to 1$^\circ$) tilt of the disk due to possible fluctuations of the pointing of the spacecraft. Another reason for implementing a toroid edge is the reduction of diffracted light by around a factor of 2 compared to a knife edge occulter \citep{Landini2011}.

The optical design of the ASPIICS telescope has been described in detail by \citet{Galy2015}. The optical scheme of ASPIICS is shown in Fig.~\ref{fig:optics}. The entrance pupil diameter is 50~mm. Internal baffles are placed between the entrance aperture and the primary objective (PO). The corona is imaged on the detector by a set of successive lenses: O1 (primary objective), O2 (field lens), and O3 (relay lense system). The calculated focal length of the telescope is 734.6~mm. 

Diffraction on the external occulter is the main source of straylight in externally occulted coronagraphs \citep{Bout2000, Aime2013, Rougeot2017, Shestov2019}. To suppress it, the image of the external occulter diffraction ring is projected by the primary objective O1 on its conjugate optical plane, where the diffraction ring is blocked by the internal occulter with the outer radius of 1.752~mm. This plane is located on the surface of the O2 field lens, and the internal occulter is slightly oversized compared to the image of the external occulter to cover the diffraction ring fully. The field lens O2 creates an image of the entrance pupil in its conjugate plane, where a circular diaphragm (the Lyot stop) is placed to block the light diffracted by the entrance pupil. The diameter of the Lyot stop is 13.85~mm, which is around 97\% of the diameter of the entrance pupil. This optical scheme has been proven successful by previous externally occulted coronagraphs, such as LASCO C2 and STEREO COR2. 

\subsubsection{Field of view and vignetting}
\label{S-vignetting}

\begin{figure}[!ht]
\centering
\includegraphics[width=0.48\textwidth]{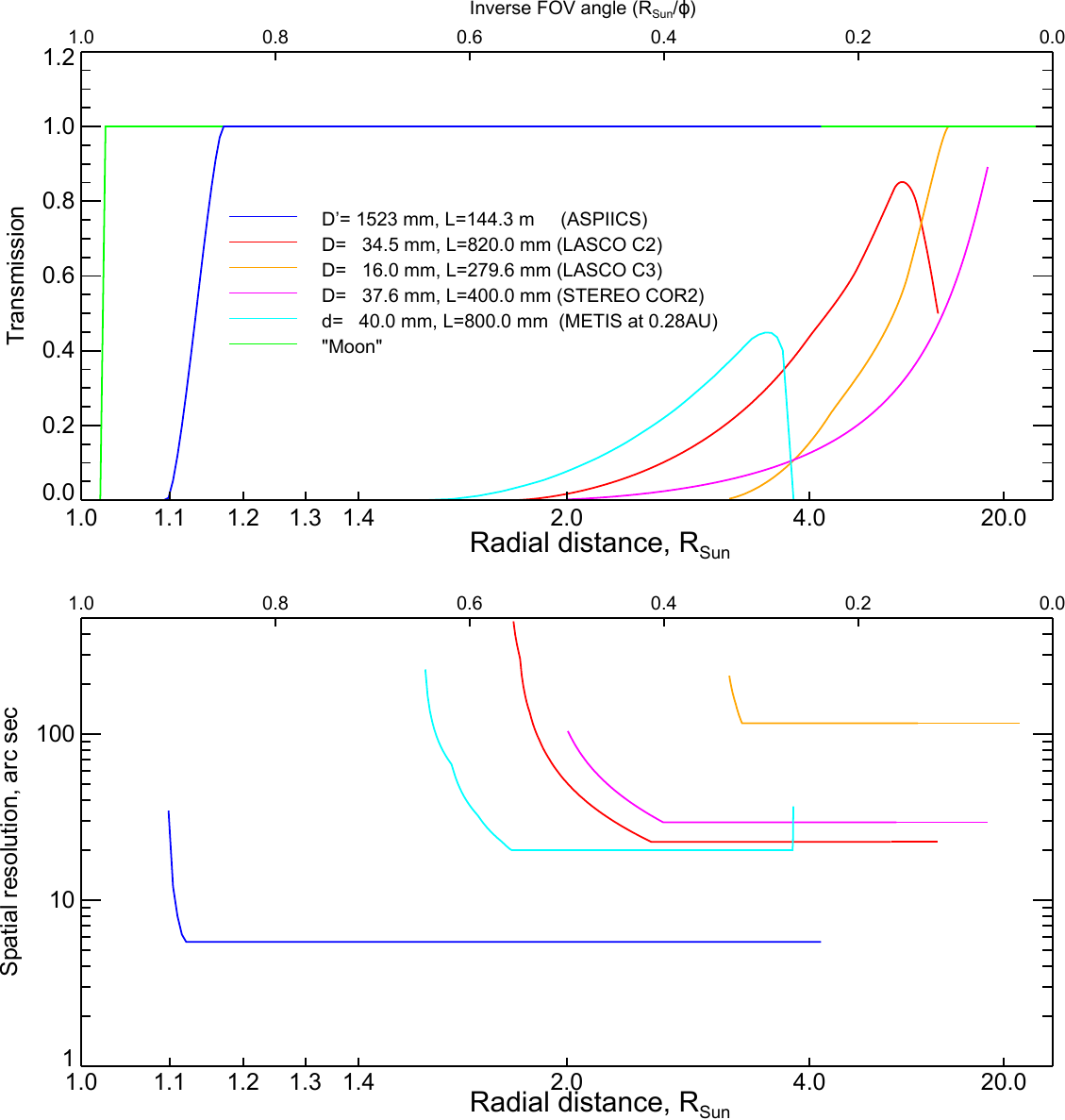}
\caption{Top panel: comparison of vignetting functions of modern externally occulted coronagraphs and the Moon. Bottom panel: comparison of spatial resolutions of modern externally occulted coronagraphs. Despite the angular resolution of Metis (20\arcsec ~in the unvignetted zone) being worse than that of ASPIICS (2.817\arcsec ~in the unvignetted zone), the linear spatial resolutions of ASPIICS and Metis are very similar (around 4000~km) when Metis observes the Sun from a vantage point near the Solar Orbiter perihelion at 0.28~au.}
\label{fig:vign}
\end{figure}

There are two types of vignetting in ASPIICS: due to external and internal occulters (narrow annular zone close to the solar limb), and due to the barrels of the spectral filters (seen at the corners of the image).

For the majority of the previous externally occulted coronagraphs, the whole field of view is partially vignetted by the external occulter, resulting in loss of optical throughput and degradation of spatial resolution. In ASPIICS, due to the large distance between the external occulter and the entrance pupil, the external occulter vignettes only a very narrow ($\sim 25$ pixels-wide) annular zone around the solar limb. The top panel of Fig.~\ref{fig:vign} shows the theoretical one-dimensional vignetting function of ASPIICS compared to the vignetting functions of other coronagraphs. As the internal and external occulters are situated in conjugated optical planes of the primary objective, the vignetting is determined by the size of the internal occulter \citep{Shestov2021}. For $r_{\rm IO} = 1.752$~mm, the field of view is unvignetted above 1.174~$R_\odot$, and below 1.099~$R_\odot$ the field of view is fully vignetted\footnote{Note that the vignetting numbers in \citet{Shestov2021} are slightly different as a different value of the internal occulter radius was assumed in that work.}. The ASPIICS vignetting profile is close to that of the Moon. Due to the eccentricity of the Earth's orbit, the apparent size of the Sun changes in the course of the year. With the size of the internal occulter being fixed, this means that the boundaries of the vignetting zone slightly change too, see Fig.~\ref{fig:ISD}. At the closest approach to the Sun, the vignetting zone starts at 1.081 ~$R_\odot$, and at the furthest approach it starts at 1.117~$R_\odot$. We emphasize that even if the variation of the edge of the vignetting zone is similar to the change of the ISD (see Fig.~\ref{fig:ISD}), the vignetting zone variation is not produced by the ISD change but only by the change of the apparent size of the Sun. 

A consequence of the vignetting profile is that the spatial resolution of ASPIICS is pixel-limited in most of the field of view (Fig.~\ref{fig:vign}). The detector pixel size is 2.817\arcsec, which gives the two-pixel spatial resolution of 5.634\arcsec. This corresponds to the linear spatial resolution of 4085~km on the Sun. In the vignetted zone, the spatial resolution is diffraction-limited and rapidly degrades near the fully vignetted zone \citep{Theys2022}.

ASPIICS has higher angular resolution than LASCO C2 and C3, Metis, and COR2 coronagraphs (Fig.~\ref{fig:vign}, bottom panel). However, it should be noted that the linear pixel-limited spatial resolution of the Metis coronagraph near the Solar Orbiter perihelion of 0.28~au (around 4300~km, two pixels) is very similar to that of ASPIICS. 

\begin{figure*}[!ht]
\centering
\includegraphics[width=1.0\textwidth]{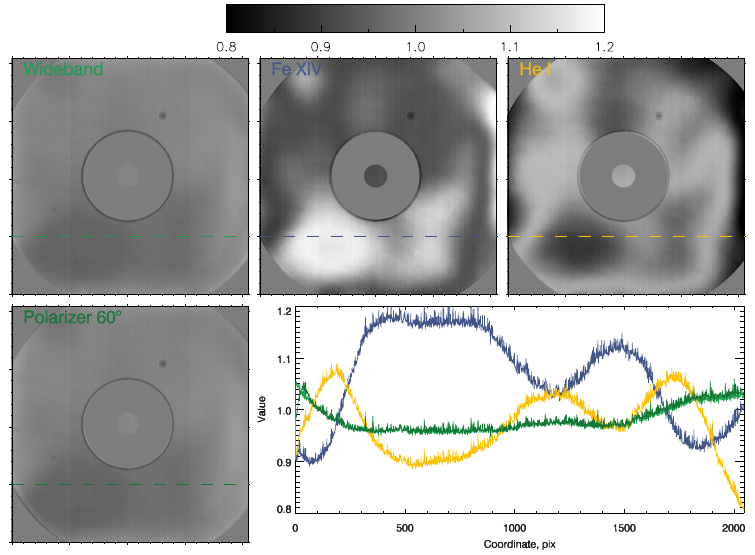}
\caption{Flat fields for different spectral channels of ASPIICS measured during the on-ground calibration. The flat fields for the other two polarized channels are very  similar to the flat fields of the Wideband and Polarizer 60$^\circ$ channels and are not shown. Note the image of the internal occulter, the hole in the internal occulter, and the vignetted corners of the field of view. The plot in the bottom right panel shows the horizontal profiles along the lines shown in the other panels.}
\label{fig:flats}
\end{figure*}

Barrels of the ASPIICS spectral filters produce vignetting in the image corners starting at $\sim 1^\circ$ (i.e. at distances above $3.75~R_\odot$ from the solar center; see Fig.~\ref{fig:flats}. Thus, the field of view of ASPIICS has a complex shape, mainly determined by the square detector with linear scale $1.6^\circ$ (from $-3$ to $3~R_\odot$), the minimal height of the unvignetted zone of $1.174~R_\odot$ determined by the internal occulter, and the maximal height of $3.75~R_\odot$ near the corners of the image.

\subsubsection{Filters and polarizers}
\label{S-filters}

\begin{figure*}[!ht]
\sidecaption
\centering
\includegraphics[width=12cm]{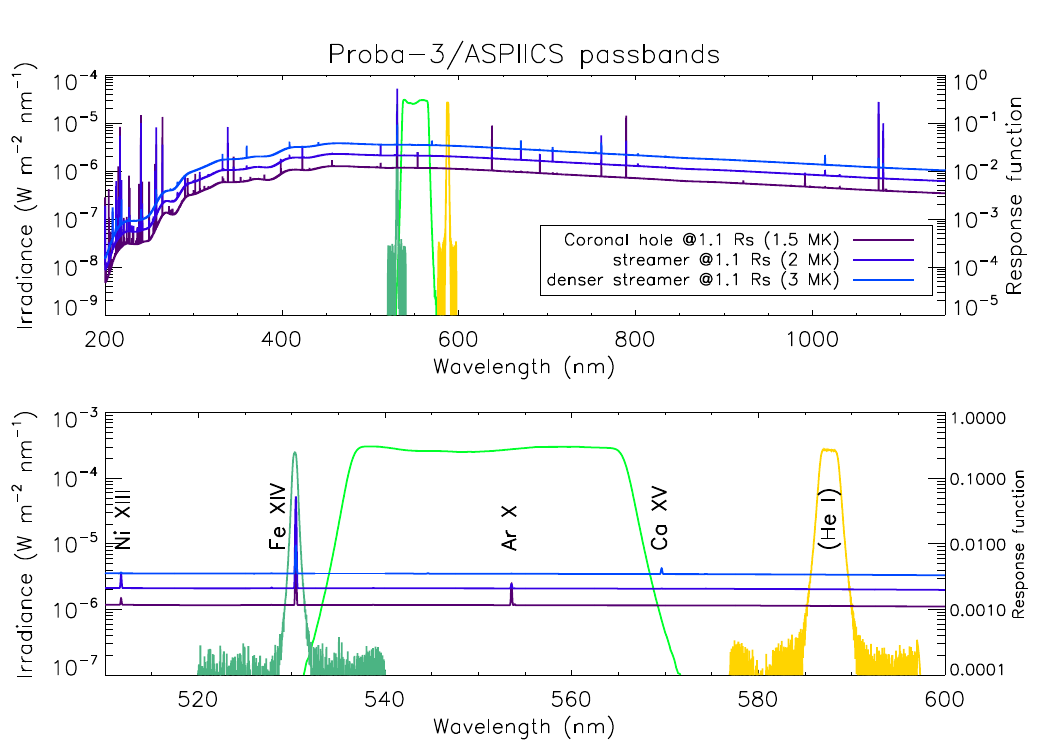}
\caption{Top panel: three spectral passbands of ASPIICS (dark green, light green, and yellow curves) plotted on top of the calculated coronal spectrum for three temperatures: 1.5~MK (corresponding to a cooler plasma, e.g. that of a coronal hole), 2~MK (streamer), and 3~MK (hot and dense streamer). Bottom panel shows a narrower spectral range with the main spectral lines marked. The dark green, green, and yellow curves in both panels show the spectral responses of the Fe~XIV, WBF, and He~I passbands of ASPIICS. }
\label{fig:passbands}
\end{figure*}

The filter wheel placed between the O3 system of relay lenses and the detector is used to select spectral and polarimetric channels, see Table~\ref{table:passbands}. The filter wheel has six slots (Fig.~\ref{fig:optics}). The main spectral channel is the wideband channel (around 300~\AA\ wide) centered around 5510~\AA\  that is dominated by the emission of the coronal continuum essentially at all temperatures of the emitting plasma\footnote{The wideband channel contains a weak Ar~X line at 5538~\AA\ \citep{DelZanna2018} that is visible only in the very hot corona (peak formation temperature of 3~MK).}. Three slots combine the same wideband channel with polarizers oriented at 0$^\circ$, 60$^\circ$, and 120$^\circ$. Two narrowband channels are also present. One is sensitive to the hot coronal emission formed at temperatures around 2~MK in the Fe~XIV line\footnote{As ASPIICS observes in vacuum, all the wavelengths are listed in vacuum.} at 5304~\AA, see e.g. \citet{DelZanna2018}. The second is sensitive to the emission of prominences in the He~I D$_3$ line at 5877~\AA. The He~I D$_3$ line is sensitive to prominence plasma under a range of conditions at temperatures between 8~10$^3$~K and 10$^5$~K \citep{Jejcic2018}. The actual spectral profiles are shown in Fig.~\ref{fig:passbands} (see also Table~\ref{table:passbands}).

\begin{table*}[!ht]
\renewcommand{\arraystretch}{1.2}
\begin{center}
\caption{Spectral passbands of ASPIICS.}
\label{table:passbands}
\begin{tabular}{llllp{4.5cm}}    
  \hline\hline                   
Short name & Passband name & Spectral peak & Passband FWHM & Main contributions \\
  \hline
WBF     & Wideband                            & 5510.6~\AA & 295.5~\AA & continuum \\
Pol 0   & Wideband + polarizer~at~0$^\circ$   & 5519.3~\AA & 296.3~\AA & polarized continuum \\
Pol 60  & Wideband + polarizer~at~60$^\circ$  & 5507.3~\AA & 295.6~\AA & polarized continuum \\
Pol 120 & Wideband + polarizer~at~120$^\circ$ & 5512.9~\AA & 295.9~\AA & polarized continuum \\
Fe XIV  & Narrowband Fe~XIV            & 5303.4 \AA & 5.4 \AA   & coronal green line \mbox{(Fe XIV, 5304~\AA)} and continuum \\
He I    & Narrowband He~I~D$_3$        & 5876.1 \AA   & 21.0 \AA  & He I D$_3$ line in prominences, continuum \\
  \hline
\end{tabular}
\end{center}
\end{table*}

\subsection{Electronics}
\label{S-electronics}

After the filter wheel, the light comes to the Focal Plane Assembly (FPA), of which the main part is the detector protected by a glass with 50\% neutral density filter (to decrease the intensity of optical ghosts). It is an Active Pixel Sensor frontside illuminated detector  that has 2048$\times$2048 square pixels, which have 10~$\mu$m size and 14 bit output. The same type of detector is mounted on the Metis coronagraph \citep{Antonucci2020} and on the Polarimetric and Helioseismic Imager \citep[PHI, see][]{Solanki2020} aboard Solar Orbiter. The detector output is transferred to the Camera Electronic Box (CEB), which controls the detector operations. 

The CEB controls the FPA by sending the required control signals, and receiving the analog images taken by the detector and digitizing them. Due to the high dynamic range of the corona and the essentially unvignetted field of view of ASPIICS, the full detector cannot be properly exposed, avoiding both saturation and underexposure. Different exposure times need to be used in different parts of the image. The complete image on the detector is divided into 32$\times$32 tiles, each having 64$\times$64 pixels. Each tile is small enough, so that a single integration time is sufficient to expose it properly. Properly exposed tiles are then selected in the CEB (see Sect.~\ref{S-operations}). The images are then sent to the Coronagraph Control Box (CCB). 

The CCB is the onboard computer that controls all activities of the instrument. It consists of the Power Conditioning Unit (PCU), Ancillary Electronics Unit (AEU), and the Data Processing Unit (DPU). The PCU provides power to all subsystems of the coronagraph. AEU operates the instrument door, the filter wheel, the Shadow Position Sensor (SPS, see Sect.~\ref{S-metrology}), and sends the commands to the CEB. The DPU interfaces with the central computer of the CSC, receives the ASPIICS data from the CEB, compresses them using an CCSDS 121.0-B-2 FPGA accelerator \citep{Kranitis2015}, and sends them to the spacecraft mass memory. All three CCB subsystems have two nearly identical units for redundancy. 

\subsection{Formation flying metrology}
\label{S-metrology}

In addition to the spacecraft formation flying metrologies described in Sect.~\ref{S-mission}, ASPIICS makes use of two additional metrologies: the Shadow Position Sensor (SPS) and the Occulter Position Sensor Emitter (OPSE). The SPS is used to determine the precise position of the ASPIICS entrance aperture with respect to the umbra produced by the external occulter. The OPSE is used to determine the position of the external occulter in the coronagraph field of view. 

\subsubsection{Shadow Position Sensor (SPS)}
\label{S-SPS}

\begin{figure}[!ht]
\centering
\includegraphics[width=0.45\textwidth]{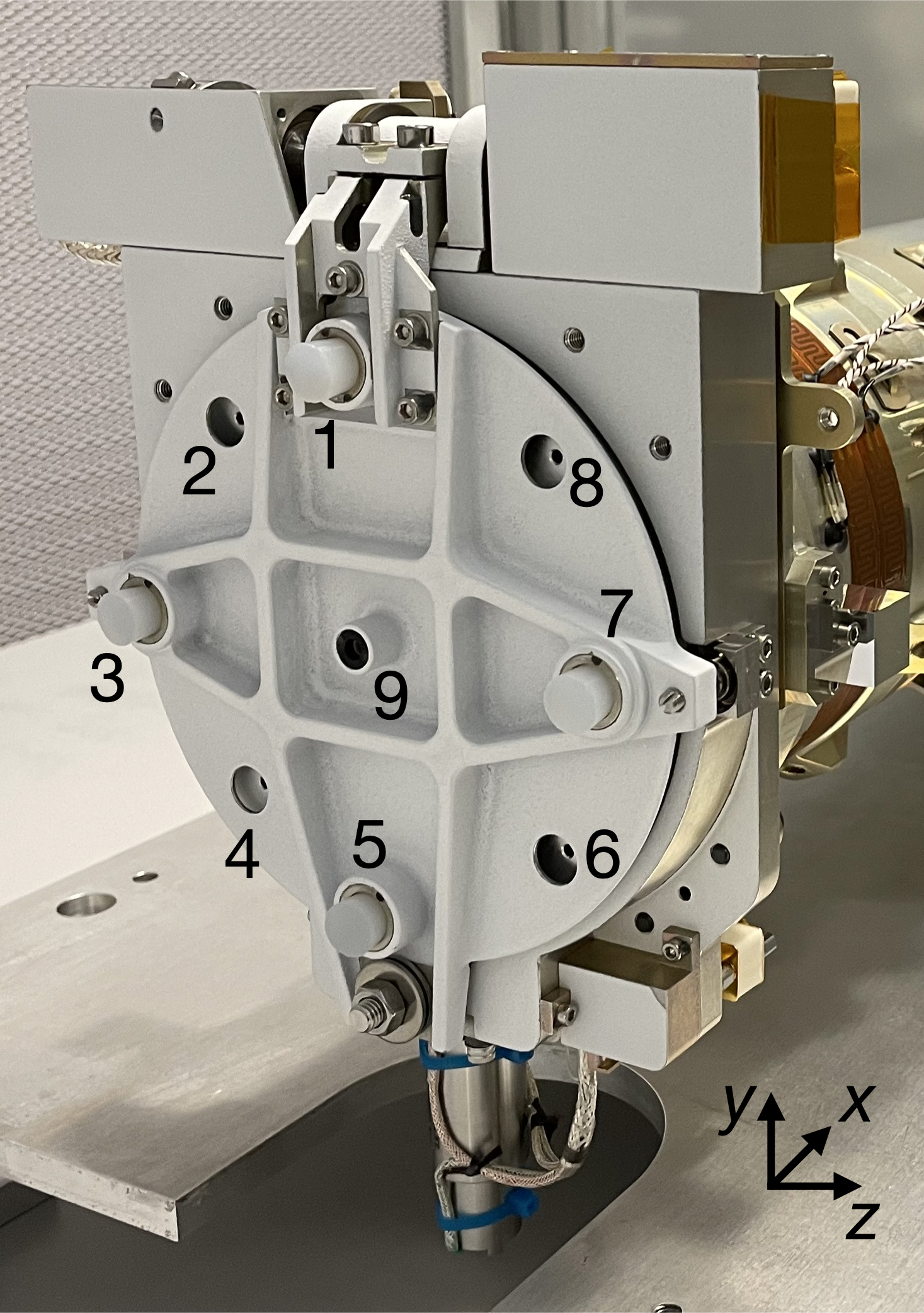}
\caption{A photo of the Front Door Assembly (FDA). The annotations highlight the locations of the SPS photo-diodes (nominal set: 1, 3, 5, 7; redundant set: 2, 4, 6, 8) and the high-density diffuser (9). The axes of the coordinate system attached to the instrument are shown, see definition in the caption of Fig.~\ref{fig:aspiics}.}
\label{fig:fda}
\end{figure}

The Shadow Position Sensor \citep[SPS, see][]{Loreggia2018, Noce2021} represents a set of eight photodiodes around the entrance aperture of ASPIICS (Fig.~\ref{fig:fda}). The difference between the readings of different photodiodes is used by an onboard algorithm within the ASPIICS software to estimate the degree of de-centering of the ASPIICS aperture with respect to the umbra produced by the external occulter. This information is used by the spacecraft and the required action is taken: either adjustment of the CSC position, or closing of the ASPIICS door (in case the de-centering is important enough to present a danger of direct sunlight entering the instrument aperture).

The photodiodes are placed in the penumbra of the external occulter, along a circle with the radius of 55~mm centered at the ASPIICS entrance aperture. Eight photodiodes are used, organized in two sets (nominal and redundant) of four photodiodes. The algorithm of the SPS is described by \citep{Casti2019}. The designed accuracy of SPS to determine the lateral displacement of the ASPIICS entrance aperture with respect to its nominal position in the umbra is 50~$\mu$m. 

\subsubsection{Occulter Position Sensor Emitter (OPSE)}
\label{S-OPSE}

The Occulter Position Sensor Emitter (OPSE) is a set of three light emitters mounted on the rear side of the OSC (the side facing the CSC), see Fig.~\ref{fig:spacecraft}. They can be directly imaged by ASPIICS due to a hole in its internal occulter (the hole diameter is 1.085~mm). The purpose of OPSE is thus to determine the position of the external occulter with respect to the coronagraph and to verify the alignment of the two spacecraft. The three emitters are arranged in a pattern of a scalene right triangle (Fig.~\ref{fig:spacecraft}, right panel), so that an image of OPSE gives unambiguous information about the position of the external occulter in the ASPIICS field of view. Information about the inter-satellite distance and the tilt of the external occulter can also be deduced. It has been shown that the straylight produced by the OPSE that reaches the ASPIICS detector through the hole in the internal occulter is negligible \citep{Galy2019}.

The information that can be derived from OPSE is not used onboard for formation flying adjustment, but is used a posteriori on the ground for the analysis of the images. 

Each of the three OPSE emitters consists of two diodes (one nominal and one redundant). The designed precision of the determination of the position of the external occulter using OPSE is 0.3~mm. 

\section{Instrument performance}
\label{S-performance}

Assembly and the first part of the on-ground calibration of ASPIICS were done at the Centre Spatial de Li\`ege, Belgium, between November 2020 and February 2021 \citep{Galy2023}. The second part of calibration was performed at OPSys facility, Turin, Italy \citep{Capobianco2019,Fineschi2025}, in September 2021.

During the assembly, the optimal focusing of the telescope was achieved by adjusting the position of the detector along its optical axis. The width of the point spread function (PSF) was measured, and its typical full width at half maximum (FWHM) was around 11.86~$\mu$m (the value was slightly different for different angles and different spectral filters), coinciding well with the FWHM of 11.16~$\mu$m of a diffraction-limited system \citep{Galy2023}. The pixel plate scale was precisely measured by rotating the telescope at angle $\pm0.7^\circ$ and measuring displacement of the collimator cross on the detector. The measured value was 2.817\arcsec\ per pixel \citep{Galy2023}. The analysis of the optical design suggested that the optical distortion of ASPIICS was negligible, and thus it was not measured during on-ground calibration.

The second part of the calibration was performed at Turin, at the OPSys facility \citep{Capobianco2019}. The following characteristics were measured: detector dark current and bias, radiometric sensitivity, flat field, detector nonlinearity, orientation of polarizers, and level of noise. In general, majority of the measured characteristics and the instrument performance coincided well with the model values \citep[see, for example,][]{Fineschi2025,Shestov2021}. For the detector bias and dark current, 2D maps have been derived from a series of measurements under dark conditions. The mean (over the field of view) value of bias was around 2050~DN, for the dark current it was around 3.3~DN~s$^{-1}$. The analysis revealed that the bias depends on temperature, with the average rate 0.6~DN~$\degr C$. The radiometric sensitivity was measured by illuminating ASPIICS with a LED-based (light-emitting diode) flat-field panel (FFP), situated few tens of centimeters in front of the entrance aperture. The signal from ASPIICS images averaged over the useful field of view was divided by the absolute brightness of the FFP in a corresponding spectral passband. In order to convert the measured values to the mean solar brightness (MSB), the solar spectrum\footnote{We used the Air Mass Zero (AM0) model, see \url{https://www.nrel.gov/grid/solar-resource/spectra}.} was multiplied with the spectral passbands of ASPIICS (see Fig.~\ref{fig:passbands}). Table~\ref{table:radiometric} summarizes the values of the radiometric sensitivity\footnote{In \citet{Fineschi2025} slightly different numbers for the radiometric sensitivity were reported, which differ mainly due to the use of the Planck function to compute the MSB values.}. We note that simplification of either of the factors -- use of a Plank function instead of AM0, or use of analytical functions (a Gaussian or a box-car) for the spectral passbands of ASPIICS -- resulted in a discrepancy of the order of only a few percent as compared to the actually used model. 

The analysis of the photon transfer curve allowed to estimate the electronic gain $g$ of the detector, which amounted to $0.10$ DN electron$^{-1}$. The photo-related signal near the saturation level was $\sim13\,000$~DN (estimated as $\sim 2^{14}$ minus the detector bias), which can be expressed as the detector full-well capacity $f_\mathrm{FWC}\sim130\times10^3$~electrons. Some drop of the photo-sensitivity started to occur near $\sim12\,000$~DN ($120\times10^3$~electron). The theoretical radiometric sensitivity calculated following the approach presented by \citet{Shestov2021} coincided with the actual measurement within 20\%, which justified the assumed values of the detector quantum efficiency of $0.65$. 

\begin{table}[!ht]
\renewcommand{\arraystretch}{1.2}
\begin{center}
\caption{Radiometric sensitivity of ASPIICS}
\label{table:radiometric}
\begin{tabular}{lp{2.5cm}p{2cm}p{2cm}}    
  \hline\hline                   
Filter & Absolute radiometric sensitivity, DN cm$^2$ sr photon$^{-1}$  & MSB value, photon s$^{-1}$ cm$^{-2}$ sr$^{-1}$ & Radiometric sensitivity, DN s$^{-1}$ MSB$^{-1}$ \\
  \hline
WBF     & $1.13\times10^{-10}$  & $2.082\times10^{20}$ & $2.36\times10^{10}$    \\
Pol 0   & $1.01\times10^{-10}$  & $2.082\times10^{20}$  & $2.11\times10^{10}$   \\
Pol 60  & $1.00\times10^{-10}$  & $2.082\times10^{20}$  & $2.09\times10^{10}$   \\
Pol 120 & $9.06\times10^{-11}$ & $2.082\times10^{20}$  & $1.89\times10^{10}$   \\
Fe XIV  & $9.71\times10^{-11}$ & $3.742\times10^{18}$ & $3.63\times10^{8}$     \\
He I    & $1.15\times10^{-10}$  & $1.548\times10^{19}$ & $1.78\times10^{9}$     \\
  \hline
\end{tabular}
\end{center}
\end{table}

Variation of the telescope sensitivity across the field of view -- the flat field --  is shown in Fig.~\ref{fig:flats} for the wideband, Fe~XIV, He~I and Polarizer 60$^\circ$ channels. The color scale of the images corresponds to the range 0.8--1.2. The flat fields are normalized in such a way that the average across the useful field of view equals unity; in the region under the IO, and in the vignetting zone it is manually set to unity. The variations of the flat field for the wideband and polarized channels of ASPIICS do not exceed a few percent, and are mainly accounted for by the slight variation of the sensitivity of the detector and the presence of dust particles inside the instrument. For example, a large dark dot seen in the flat fields (to the top right of the internal occulter) is most probably due to the presence of a small dust particle somewhere in the vicinity of the detector (as it is not focused), e.g. on the 50\% neutral density filter. The flat fields drop by around 10\% in these areas.

The detector nonlinearity is attributed to the nonlinear conversion of the photo-electrons into the DN (Digital Numbers). The nonlinearity function, defined as the ratio of the measured and the linear response minus one, did not exceed 1\% across the most of the dynamic range. The random noise, which is a sum of the detector read-out noise, photon and dark current shot noises \citep{Shestov2021}, was derived from a series of images taken with different exposure times using the photon transfer curve analysis. The derived values range from $\sim 7$~DN for the minimal exposure (thus being a read-out noise) to $\sim 36$~DN for the exposures approaching saturation of the detector, which is in agreement with the estimated detector gain $g = 0.10$~DN~electron$^{-1}$ and full-well capacity $f_\mathrm{FWC}\sim130\times10^3$~electrons.

Measurements of the polarizers orientation were performed with an additional pre-polarizer placed between the FFP and the telescope. The pre-polarizer was rotated in small steps, and during subsequent data analysis the parameters of ASPIICS polarizers were calculated on the per-pixel basis. These data were interpreted either as a variation of the polarizer orientation across the FOV (Shestov et al., in prep.), or as a demodulation tensor for every pixel across the FOV \citep{Casti2018,Fineschi2025}, with both approaches giving consistent results. The averaged angles of the polarizers amounted to 5.3, 65.6 and 125.4$^\circ$ with a variation up to $\pm0.5^\circ$ across the field of view. 

Due to the difficulties related to the measurement of the diffracted light in ASPIICS, it was not measured on-ground. Instead, we rely on the theoretical models \citep{Aime2013, Rougeot2017, Shestov2018}. If these models are combined with provisional ASPIICS characteristics and simplified radiometric model \citep{Shestov2021}, they show that diffracted light is two orders of magnitude weaker than the coronal signal. Such a small contribution of the diffracted light has been confirmed during the commissioning phase of ASPIICS \citep{Shestov2026}. 

\section{Operational concept}
\label{S-operations}

\subsection{Tile maps}
\label{S-tiles}

\begin{figure}[!ht]
\centering
\includegraphics[width=0.48\textwidth]{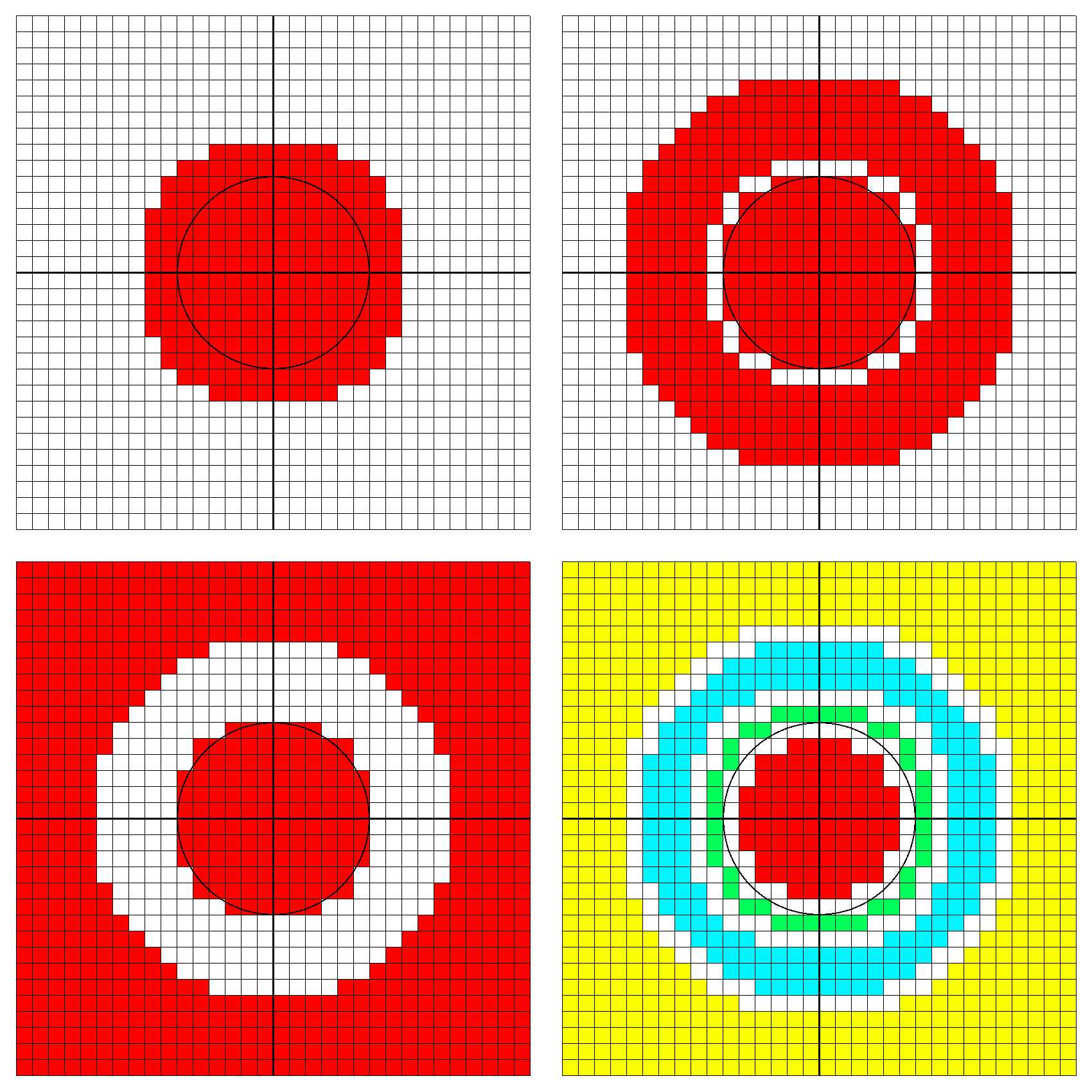}
\caption{ASPIICS tile maps. Top left, top right, and bottom left panels show examples of 3 tile maps used to recover the
characteristic dynamic range of the coronal brightness in the wideband filter images taken with different exposure times. In these examples, 224, 444, and 756 tiles are retained, shown in red, for Map1, Map2, and Map3, which register the inner, middle and outer parts of the vield of view, respectively. Each tile map is compound of 1024 tiles, 64$\times$64 pixels each. Central tiles are registered in each image to reveal the LEDs of the OPSE. The bottom right panel summarises the tile map regions, with white tiles corresponding to the overlapping tiles which are registered twice in two different exposures to ease their stitching process on the ground.}
\label{fig:tiles}
\end{figure}

In order to match the very large dynamic range of brightness in the corona, images can be recombined onground from up to 3 consecutive frames taken in the same passband with different exposure times. To avoid downlinking underexposed or overexposed tiles in a frame to the ground, two approaches are combined.

First, predefined tile maps specify which tiles must be kept for a given exposure. These tile maps are compound of 1024 tiles, 64$\times$64 pixels each. The tile maps define three overlapping nearly circular concentric domains with different distances from the solar centre (Fig.~\ref{fig:tiles}). The use of tile maps with a reduced number of tiles accelerates the transfer of the recorded images from the camera to the instrument computer (the transfer speed is constant, amounting to 192 tiles per second). Together with reducing the size of the stored data, it therefore favors the achievement of faster observation cadences. 

Second, tiles can be discarded depending on the quality flag. The quality flag is applied to each individual tile that is selected by the tile map in the previous step. It introduces no significant delay in the processing of the data, but does not accelerate this processing either (the transfer speed from the camera to the instrument computer being the limiting factor). The quality flag is defined by a set of 3 parameters: the value in DN in the raw image above which a pixel should be deemed not underexposed (underexposure threshold, for example above the level of expected dark current), the value in DN above which the pixel is considered saturated or too close to the saturation level (overexposure threshold), and the minimum number of pixels in the tile that should be present between the previous two thresholds (quality threshold). If the number of "correctly exposed" pixels is larger than the quality threshold, then the tile is transferred to the instrument computer. Used in combination with "conservative" tile maps (with a lot of overlapping between the tile maps of different exposure times, see Fig.~\ref{fig:tiles}), the quality flag automatically adjusts to the generally asymmetric corona and further saves storage space in the onboard mass memory. The extreme case is that of the prominences in He I D$_3$ images, where only a few tiles may contain significant signal with the exposure times that are typically used, at the location of prominences.

\subsection{Observing programs}
\label{S-programs}

The camera can take up to 3 successive exposures in burst mode, with minimal delay between them (exposure time and detector readout). They are meant to capture the high dynamic range of the corona with, typically, exposure times differing by a factor 10 (e.g. 0.1~s, 1.0~s, and 10~s in the wideband). 
Together, they form one acquisition obtained with the same filter. Each image/exposure can be associated to a different tile map, and each acquisition can be associated with a different set of quality flag parameters. Between each acquisition, the filter wheel can optionally be rotated to change the filter.  

Acquisitions can be grouped in a sequence of up to six (corresponding to the 6 different ASPIICS filters), and repeated a certain number of times (a cycle), to build the observing programs. 

The observing programs are fully flexible, but a minimal set of 4 programs is intended to match the ASPIICS science objectives: 

1. The Full Set program corresponds to a sequence of 6 acquisitions, one for each filter, with 2 or 3 exposure times for each, to capture the dynamic range of the coronal signal at the best signal-to-noise ratio. It takes a bit less than 4 minutes to complete and is typically intended to observe the quiescent corona (as it includes long exposure times), usually at the beginning and end of each 6-hour coronagraphy interval. 

2. The Synoptic program consists of a Full Set followed by 54 acquisitions in the wideband filter at a cadence of 60~s (or alternatively, 106 acquisitions at a cadence of 30~s). It lasts 1 hour and can be repeated 6 times to fill in one coronagraphy interval. 

3. The Waves programs consist of fast cadence observations in one filter (usually wideband), typically at 2~s, 4~s or 15~s cadence, depending on the number of exposures and number of tiles in the tile maps (adapting the field of view accordingly for the desired signal-to-noise ratio). 

4. The CME-Watch program consists of the wideband and narrowband filters observed every 60~s, replaced by acquisitions in the 3 polarized filters every 5 minutes. 

The Waves and CME-Watch programs produce much more data volume in 6 hours than can be downloaded in one day, so that one needs to alternate them with programs producing less data, or use selective downlink. The science data compression is made using an CCSDS 121.0-B-2 FPGA accelerator \citep{Kranitis2015}. The baseline compression is lossless, although the use of a lossy compression is also possible.

\subsection{Data products}
\label{S-products}

\begin{table}[t]
\renewcommand{\arraystretch}{1.2}
\begin{center}
\caption{ASPIICS data products.}
\label{table:dataproducts}
\begin{tabular}{lll}    
  \hline\hline                   
Data level & Data products & Units\tablefootmark{a}  \\
  \hline
 Level-1 & Wideband  & DN \\
         & Polarizer 0$^\circ$ & DN \\
         & Polarizer 60$^\circ$ & DN \\
         & Polarizer 120$^\circ$ & DN \\
         & Fe XIV passband & DN \\
         & He I D$_3$ passband & DN \\
  \hline
 Level-2 & Wideband  & MSB \\
         & Polarizer 0$^\circ$ & MSB \\
         & Polarizer 60$^\circ$ & MSB \\
         & Polarizer 120$^\circ$ & MSB \\
         & Fe XIV passband & MSB \\
         & He I D$_3$ passband & MSB \\
  \hline
 Level-3 &  Total brightness & MSB \\
         &  Polarized brightness & MSB \\
         &  Fe XIV line emissivity & photons s$^{-1}$ cm$^{-2}$ sr$^{-1}$ \\
         &  He I D$_3$ line emissivity & photons s$^{-1}$ cm$^{-2}$ sr$^{-1}$ \\
  \hline
\end{tabular}
\tablefoottext{a}{DN: Data Number; MSB: Mean Solar Brightness. }
\end{center}
\end{table}

The ASPIICS data products are stored and shared as FITS (Flexible Image Transport System) files in 3 different levels of processing. 

Level-1 is the reference dataset that is maximally documented by all available metadata. It keeps the original image data untouched and is meant for users with an interest in instrument and calibration issues. 

Level-2 is the reference science dataset that is calibrated to the best of the knowledge of the instrument team. It serves most of the needs of the external science community by abstracting as much as possible the instrument peculiarities. At Level-2, each file only contains the data of 1 exposure. Although they contain the full metadata about the position of the Sun in the image and the attitude of the spacecraft, no recentering or rotation is applied to limit interpolation effects.

Level-3 is the dataset that stores derived data products that are optimized for particular scientific purposes. A derived Level‐3 image is constructed out of several Level‐2 images, at the sacrifice of temporal resolution, and synthetic metadata are added. All exposures of an acquisition are merged to provide the full FOV images with maximal signal-to-noise ratio everywhere. The center of the Sun is set at the center of the image and the image is rotated to have the Solar North up. The continuum component is removed from the narrow passbands, using the information about continuum acquired in the wideband image that is nearest in time.  

Table~\ref{table:dataproducts} summarizes the ASPIICS data products\footnote{In addition, there is a Level-0 dataset that contains the science data only with the metadata traveling with the telemetry. It is used as a basis for potential reprocessing and is not publicly available.}. The Level‐3 data will be also provided as PNG (Portable Network Graphics) images and MPEG (Moving Picture Experts Group) movies. All the ASPIICS data are freely available\footnote{At the website of the Proba-3/ASPIICS Science Center, see \url{https://www.sidc.be/proba-3/aspiics-data}} in the framework of the open data policy.

\section{Summary}
\label{S-summary}

\mbox{Proba-3} is a mission driven by both science and technology. It is first of all a technology demonstration mission: \mbox{Proba-3} tests precise formation flying techniques and technologies that can be used by future missions. \mbox{Proba-3} is also a Mission of Opportunity in the ESA Science Programme, as the \mbox{Proba-3/ASPIICS} capabilities represent a significant advance from previous, current, and planned solar coronagraphs. Due to the unique separation between the telescope and the external occulter (around 144~m), ASPIICS is able to observe the inner corona as close to the solar centre as 1.099~$R_\odot$ in low straylight conditions. ASPIICS can therefore fill the gap between the low corona (typically observed by EUV imagers like SDO/AIA) and the high corona (typically observed by externally occulted coronagraphs like SOHO/LASCO C2), where observations are difficult. ASPIICS observations will be crucial for solving several outstanding problems in solar physics, such as structure of the magnetized solar corona, sources of the slow solar wind, 
onset and early acceleration of coronal mass ejections. ASPIICS heralds the new generation of solar space coronagraphs.

\begin{acknowledgements}
      {The ASPIICS project is developed under the ESA’s General Support Technology Programme (GSTP) and the ESA’s PRODEX Programme thanks to the contributions of Belgium, Poland, Romania, Italy, Ireland, Greece, and the Czech Republic. The ROB team thanks the Belgian Federal Science Policy Office (BELSPO) for the provision of financial support in the framework of the PRODEX Programme of the European Space Agency (ESA) under contract numbers 4000117262, 4000134474, 4000136424, 4000145189, and 4000147286. S.J. acknowledges the support from the Slovenian Research Agency No. P1-0188. S.G., P.H., and S.J. acknowledge the support from grant 25-18282S of the Czech Science Foundation and from the project RVO:67985815 of the Astronomical Institute of the Czech Academy of Sciences. P.H. was partially supported by the program ``Excellence Initiative - Research University'' for the years 2020--2026 for the University of Wroc\l{}aw, project No. BPIDUB.4610.96.2021.KG. U.B.S. and M.S. were supported by grant No. 2024/55/B/ST9/03199 of the National Science Centre, Poland. We acknowledge enthusiastic support by Jean Arnaud (1945--2010) and Rainer Schwenn (1941--2017) during the early stages of the ASPIICS project. We dedicate this paper to the memory of Serge Koutchmy (1940--2023), solar physicist and passionate eclipse observer, who not only contributed to the scientific preparation and development of Proba-3/ASPIICS, but was also a friend, an inspiration, and a role model to many of us.}
\end{acknowledgements}

\bibliographystyle{aa}
\bibliography{bibliography.bib}

\end{document}